\documentclass[aps,prl,twocolumn,superscriptaddress,longbibliography]{revtex4-1}
\usepackage{amsmath}
\usepackage{hyperref}
\hypersetup{
	colorlinks=true,
	linkcolor=blue,
	filecolor=blue,
	citecolor = blue,      
	urlcolor=blue,
}

\usepackage{graphicx}% Include figure files
\usepackage{dcolumn}% Align table columns on decimal point
\usepackage{bm}% bold math
\usepackage{natbib}
\usepackage[utf8]{inputenc}
\usepackage{units}
\usepackage{amsmath}
\usepackage{amssymb}
\usepackage{graphicx}
\usepackage{bm}
\usepackage{multirow,microtype,color,relsize,ulem}

\usepackage[utf8]{inputenc}
\usepackage[T1]{fontenc}
\usepackage{mathptmx}
\usepackage{amsfonts}
\usepackage{amsmath}

\begin{document}

\title{Generating multiparticle entangled states by self-organization of driven ultracold atoms}
\author{Ivor Kre\v{s}i\'{c}} \email{ivor.kresic@tuwien.ac.at} 
\affiliation{Institute for Theoretical Physics, Vienna University of Technology (TU Wien), Vienna, A–1040, Austria}
\affiliation{Centre for Advanced Laser Techniques, Institute of Physics, Bijeni\v{c}ka cesta 46, 10000, Zagreb, Croatia}
\author{Gordon R. M. Robb} 
\affiliation{SUPA and Department of Physics, University of Strathclyde, Glasgow G4 0NG, Scotland, UK}
\author{Gian-Luca Oppo} 
\affiliation{SUPA and Department of Physics, University of Strathclyde, Glasgow G4 0NG, Scotland, UK}
\author{Thorsten Ackemann} 
\affiliation{SUPA and Department of Physics, University of Strathclyde, Glasgow G4 0NG, Scotland, UK}

\date{\today}% It is always \today, today,

\begin{abstract}
We describe a mechanism for guiding the dynamical evolution of ultracold atomic motional degrees of freedom toward multiparticle entangled Dicke-squeezed states, via nonlinear self-organization under external driving. Two examples of many-body models are investigated. In the first model, the external drive is a temporally oscillating magnetic field leading to self-organization by interatomic scattering. In the second model, the drive is a pump laser leading to transverse self-organization by photon-atom scattering in a ring cavity. We numerically demonstrate the generation of multiparticle entangled states of atomic motion and discuss prospective experimental realizations of the models. For the cavity case, the calculations with adiabatically eliminated photonic sidebands show significant momentum entanglement generation can occur even in the “bad cavity” regime. The results highlight the potential for using self-organization of atomic motion in quantum technological applications.

\end{abstract}

\maketitle

%\textit{Introduction}\label{sec:intro}.
The study of self-organization of ultracold atoms is a well-established research direction, with many notable experimental and theoretical results \cite{ritsch_cold_2013,mivehvar_cavity_2021}. Following the pioneering works on self-organization of cold \cite{domokos_collective_2002,black_observation_2003} and ultracold  \cite{baumann_dicke_2010,nagy_dicke-model_2010} atoms coupled to a single longitudinal mode of a Fabry-Perot cavity, the multimode aspects of optomechanical self-organization in cold and ultracold atoms have recently started to generate significant interest \cite{gopalakrishnan_emergent_2009,gopalakrishnan_atom-light_2010,greenberg_bunching-induced_2011,tesio_theory_2014,tesio_kinetic_2014,labeyrie_optomechanical_2014,robb_quantum_2015,schmittberger_spontaneous_2016,ostermann_spontaneous_2016,ballantine_meissner-like_2017,leonard_supersolid_2017,kollar_supermode-density-wave-polariton_2017,zhang_long-range_2018,vaidya_tunable-range_2018,guo_sign-changing_2019,guo_emergent_2019,schuster2020,baio_multiple_2021,guo_optical_2021,ackemann_self-organization_2021}. In parallel to the work on optomechanical self-organization, there has been great progress in studying the atomic self-organization arising due to atom-atom interactions being modulated by an external B-field \cite{engels2007,kronjager2010,pollack2010,nguyen2019,zhang_pattern_2020}.

Although the majority of these works have studied the nonequilibrium phase diagrams in the mean field limit, where quantum correlations can be neglected, a number of works have shown that the quantum nature of light and matter can play an important role for self-organization \cite{lugiato_quantum_1992,grynberg_quantum_1993,gatti_spatial_2001,cola_robust_2004,mekhov_probing_2007,nagy_nonlinear_2009,nagy_dicke-model_2010,nagy_critical_2011,elliott_multipartite_2015,ostermann_unraveling_2020,ivanov_feedback-induced_2020}.

Quantum correlated squeezed and entangled states can be used for quantum enhanced measurements, which go beyond classical metrology \cite{giovannetti_quantum_2006,clerk_introduction_2010}. In this context, squeezing and entanglement of the internal atomic degrees of freedom~\cite{duan2000,pu2000,sorensen_many-particle_2001,gross_nonlinear_2010,leroux_implementation_2010,schleier-smith_squeezing_2010,gross_spin_2012,hosten_measurement_2016,luo2017,pezze_quantum_2018,davis19,anjun2021}, but recently also the external ones~\cite{salvi2018,gietka2019,shankar2019,anders2021,greve2022entanglement}, have been recognized as attractive tools for metrological applications. 

Here, we focus on external degrees of freedom and study the spontaneous generation of multiparticle entangled (also called Dicke squeezed) states in the atomic motion \cite{dunningham2002,lucke_twin_2011,bucker_twin-atom_2011,duan_entanglement_2011,zhang_quantum_2014,lucke_detecting_2014,pezze_quantum_2018}, by self-organization of an externally driven Bose-Einstein condensate (BEC). The mechanism of this phenomenon is four-wave mixing \cite{deng1999,vogels2002,gemelke2005,campbell2006,perrin2007,dall2009,jaskula2010} of zero-order modes with the spontaneously generated transverse sidebands, and we demonstrate the effect numerically for two different models. In the first model, the self-organization arises due to an external periodic modulation in interatomic scattering, driven by a temporally oscillating B-field. In the second model, it arises from laser driving in a ring cavity. A confinement along the $y$ and $z$ axes allows us to restrict the analysis to 1D structures in an elongated cloud.

\begin{figure}[!t]
\centering
\includegraphics[clip,width=\columnwidth]{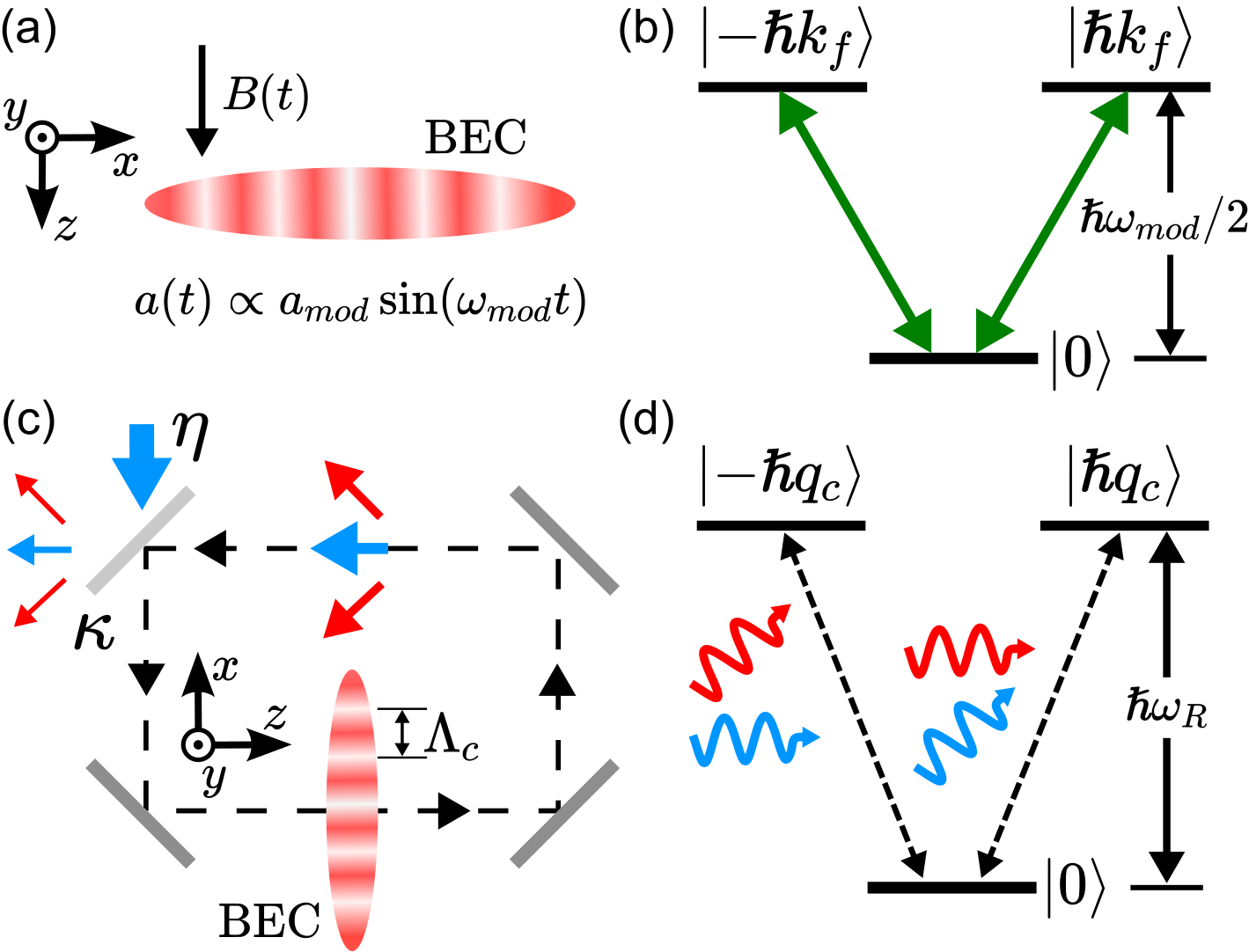}
\caption{Nonlinear self-organization via four-wave mixing of momentum modes in a driven BEC. (a) The self-organization along the $x$ axis can be driven by applying a B-field $B(t)$ oscillating near a Feshbach resonance, leading to an oscillating scattering length $a(t)$ \cite{zhang_pattern_2020}. (b) Absorption of a quantum of energy $\hbar\omega_{mod}$ leads here to a momentum-conserving scattering of two atoms with zero transverse momentum into modes with transverse momenta $\pm\hbar k_f$, where $k_f=\sqrt{m\omega_{mod}/\hbar}$. (c) Transverse self-organization with spatial period $\Lambda_c$ for laser pumped atoms in a ring cavity ($\eta$ - pump rate) with photon leakage rate $\kappa$. (d) In this system, an atomic sideband with transverse momentum $-\hbar q_c$ ($q_c=2\pi/\Lambda_c$) is excited by scattering of an on-axis photon into a vacuum mode with transverse wavenumber $+ q_c$. A correlated atom with transverse momentum $+\hbar q_c$ can then be created if this sideband photon is scattered back into the on-axis mode, with a converse process for the $- q_c$ photon sideband (not shown).}\label{Fig:1}
\end{figure}

\textit{Driving by B-field modulation.} In the first model, shown in Fig. \ref{Fig:1}a), a spatially homogeneous and temporally oscillating B-field in the $z$-direction is modulated near a Feshbach resonance of the atoms. The B-field sinusoidally modulates the atomic s-wave scattering length with frequency $\omega_{mod}$ and amplitude $a_{mod}$, thus driving the pattern formation \cite{zhang_pattern_2020}. For a collisionally thin medium, this four-wave mixing process is described by the three-mode Hamiltonian:
\begin{align}\label{eq:ham_sc}
H_B=i\hbar g_{mod} b_+^\dagger b_-^\dagger b_0 b_0+\mbox{H.c.},
\end{align}
where $g_{mod}=2\pi\hbar a_{mod}/mV$, $m$ is the atom mass, $V$ is the volume of the condensate, and $b_0,\:b_\pm$ are the bosonic momentum annihilation operators for the transverse modes of momenta $0,\:\pm \hbar k_f$, where $k_f=\sqrt{m\omega_{mod}/\hbar}$. The mechanism of pattern formation in this system is illustrated in Fig.~\ref{Fig:1}b). When a pair of atoms with zero transverse momenta absorbs a quantum of energy $\hbar \omega_{mod}$, they scatter into modes with opposite momenta and kinetic energies $\hbar \omega_{mod}/2$. This momentum conserving process leads in the semiclassical picture to a formation of stripe patterns in the atomic density.

Note that the Hamiltonian $H_B$ describes the same physics for atoms as the two-photon formalism of \cite{louisell61,caves84,yurke86,schnabel17} does for photon pairs generated in optical parametric amplifiers. Moreover, $H_B$ is the atomic momentum equivalent of the Hamiltonian used in SU(1,1) spin-1 atom interferometry \cite{linnemann16,szigeti17}. 

%Finally, a similar model to $H_B$ was also initially studied in laser driven materials with cavity feedback, where the medium's degrees of freedom were eliminated and only the photonic modes were relevant \cite{lugiato_quantum_1992,grynberg_quantum_1993,gatti_spatial_2001}.

\textit{Laser driving with ring cavity feedback.} The second model we study is based on laser driving of a BEC in a ring cavity, as illustrated in Fig.~\ref{Fig:1}c). This novel setup is inspired by earlier theoretical proposals \cite{lugiato_spatial_1987,lugiato_stationary_1988,lugiato_quantum_1992,grynberg_quantum_1993,gatti_spatial_2001,tesio_spontaneous_2012}. Such transverse self-organization has been observed in nearly degenerate Fabry-Perot cavity experiments with vertical-cavity regenerative amplifiers \cite{ackemann_spatial_2000} and photorefractive crystals \cite{esteban-martin_experimental_2004,esteban05}. For a BEC in a cavity, in addition to the three atomic motional modes with annihilation operators $b_0,\:b_\pm$ for transverse momenta $0,\:\pm\hbar q_c$ (with $q_c=2\pi/\Lambda_c$ for the pattern lengthscale $\Lambda_c$), we have also the corresponding intracavity photonic modes with annihilation operators $a_0,\:a_\pm$. In this system, a continuous-wave (cw) laser of frequency $\omega$ drives the zero-order cavity mode of frequency $\omega_0$ with pump rate $\eta$. Above some critical pump level $\eta_c$, this leads to spontaneous generation of sideband modes with frequency $\omega_0'$, concurrently with atomic momentum sideband modes. The effective Hamiltonian of the problem is given by $H_{cav}=H_0+H_{FWM}^{(1)}+H_{FWM}^{(2)}$, with:
\begin{equation}\label{eq:h0}
\begin{aligned}
%\begin{split}
H_0 &= -\hbar\bar{\Delta}_cn_0-\hbar\bar{\Delta}_c'(n_++n_-)\\
+&\hbar\omega_R(N_++N_-)+i\hbar \eta( a_0^\dagger- a_0),
%\end{split}
\end{aligned}
\end{equation}
and the four-wave mixing terms:
\begin{align}\label{eq:fwmterms_1}
\begin{split}
H_{FWM}^{(1)}& = \hbar U_0 [(a_+^\dagger b_-^\dagger + a_-^\dagger b_+^\dagger)a_0 b_0 +\mbox{H.c.}] \\
&+\hbar U_0 [a_0^\dagger(b_+^\dagger a_+ +b_-^\dagger a_-)b_0 +\mbox{H.c.}], 
\end{split} \\
H_{FWM}^{(2)}& = \hbar U_0(a_+^\dagger a_- b_-^\dagger
b_++\mbox{H.c.})\label{eq:fwmterms_2},
\end{align}
where $\bar{\Delta}_c=\omega-\omega_0-NU_0$, $\bar{\Delta}_c'=\omega-\omega_0'-NU_0$ are the effective pump detunings from the on-axis and sideband cavity modes, respectively,
%(we put $\Delta_c'=\Delta_c$), 
$n_{0}=a_{0}^\dagger a_{0},\: n_{\pm}=a_{\pm}^\dagger a_{\pm}$, $N_{0}=b_{0}^\dagger b_{0},\:N_{\pm}=b_{\pm}^\dagger b_{\pm}$,
%$\Delta_c'=\omega-\omega_0'$ are the laser-cavity detunings for the on-axis and sideband modes, respectively, 
$U_0=g_0^2/\Delta_a$ is the single atom light shift, $\Delta_a=\omega-\omega_a$ is the laser-atom detuning of the two-level optical transition, $g_0$ is the atom-cavity coupling strength, and $\hbar\omega_{R}=(\hbar q_c)^2/2m$ is the transverse recoil energy. Following \cite{nagy_dicke-model_2010,baumann_dicke_2010}, we here concentrate on the system dynamics for $\bar{\Delta}_{c}'<0$, where $\omega_0'$ (i.e. $\Lambda_c$) is tunable via Fourier filtering of the intracavity light \cite{jensen_manipulation_1998}.

Generation of correlated atom pairs via the four-wave mixing terms of $H_{FWM}^{(1)}$ can be explained by a scattering process illustrated in Fig. \ref{Fig:1}d). A photon scattered from the coherent on-axis mode into the initially empty (vacuum) sideband mode $a_\pm$, via $a_\pm^\dagger  a_0 b_\mp^\dagger b_0$, can be scattered back into the on-axis mode via $a_0^\dagger a_\pm b_\pm^\dagger b_0$. This momentum conserving process creates a correlated pair of indistinguishable atoms with opposite momenta, which can lead to multiparticle entanglement during transient dynamics in a dissipative cavity, as numerically demonstrated in this Letter.

The process can also be described by the atom-only Hamiltonian $H_{ad}$, for which the photonic modes are adiabatically eliminated (valid for $|\bar{\Delta}_c|,\: |\bar{\Delta}_c'|\gg \omega_R$), given by $H_{ad}=H_{pair}+H_{ad}'$, where:
\begin{align}\label{eq:adelim}
H_{pair}& = -\hbar g_{cav}b_+^\dagger b_-^\dagger b_0 b_0+\mbox{H.c.},% \\
%\begin{split}
%H_{ac}'&=\hbar\omega_R(N_++N_-) \\
% -&\hbar \omega_R\left[  (2N_0-1)(N_++N_-)-2N_0\right]\frac{1}{4N}\frac{\eta^2}{\eta_c^2},
%\end{split}
\end{align}
with parameter $g_{cav}=2U_0^2\eta_{eff}^2|\bar{\Delta}_c'|/[(\bar{\Delta}_c'^2+\kappa^2)(\bar{\Delta}_c^2+\kappa^2)]$ (where $\bar{\Delta}_c'<0$). In deriving $H_{ad}$ we have neglected $H_{FWM}^{(2)}$, the influence of photonic sideband creation/annihilation on the coherent field in the on-axis mode, along with the influence of photonic sideband quantum noise on the atomic motion. The relationship of $H_B$ and $H_{cav}$ becomes now more apparent. Indeed, $H_{pair}$ has the form equivalent to $H_B$, while $H_{ad}'$ describes the energy shifts related to self-organization in the laser driven system ($H_{ad}'$ and details of the derivation are given in \cite{kresic_supplemental_nodate}).
%Generation of transverse sidebands via the four-wave mixing terms of $H_{cav}$ can be explained by the momentum conserving processes illustrated in Fig. \ref{Fig:1}d). Two opposite atomic momentum sidebands are created by the term $a_+^\dagger a_0b_{-}^\dagger b_0$ in $H_{FWM}^{(1)}$ by scattering an on-axis photon into the $+q_c$ mode, which excites an atomic state with transverse momentum $-\hbar q_c$, and the term $a_{-}^\dagger a_0b_{+}^\dagger b_0$, which scatters an on-axis photon into the $-q_c$ mode and excites an atomic state with transverse momentum $\hbar q_c$. 

%In the semiclassical picture, the spontaneously generated transverse lattice potential attracts atoms towards the maxima of the light intensity for $U_0<0$, or their minima for $U_0>0$. The ordering of atoms in turn increases the light diffraction and thus also the depth of the lattice, which turns into a runaway process when the pump rate $\eta$ is sufficiently strong to compensate for the kinetic energy cost and dissipation. 

\textit{Dicke squeezed states in transverse atomic momentum}\label{sec:entang}. Our aim is to study the phenomenon of Dicke squeezing and the associated multiparticle (many-atom) entanglement for the Hamiltonians $H_B$, $H_{ad}$ and $H_{cav}$. To this end we define the sideband operators: $\delta n=n_+-n_-$, $\delta N=N_+-N_-$, along with: $J_x=(b_+^\dagger b_-+b_-^\dagger b_+)/2,\; J_y=(b_+^\dagger b_--b_-^\dagger b_+)/2i,\; J_z=\delta N/2$, which are analogous to Schwinger's angular momentum operators~\cite{schwinger51}, with $J_{eff}^2=J_x^2+J_y^2$. 

\begin{figure}[!t]
\centering
\includegraphics[clip,width=\columnwidth]{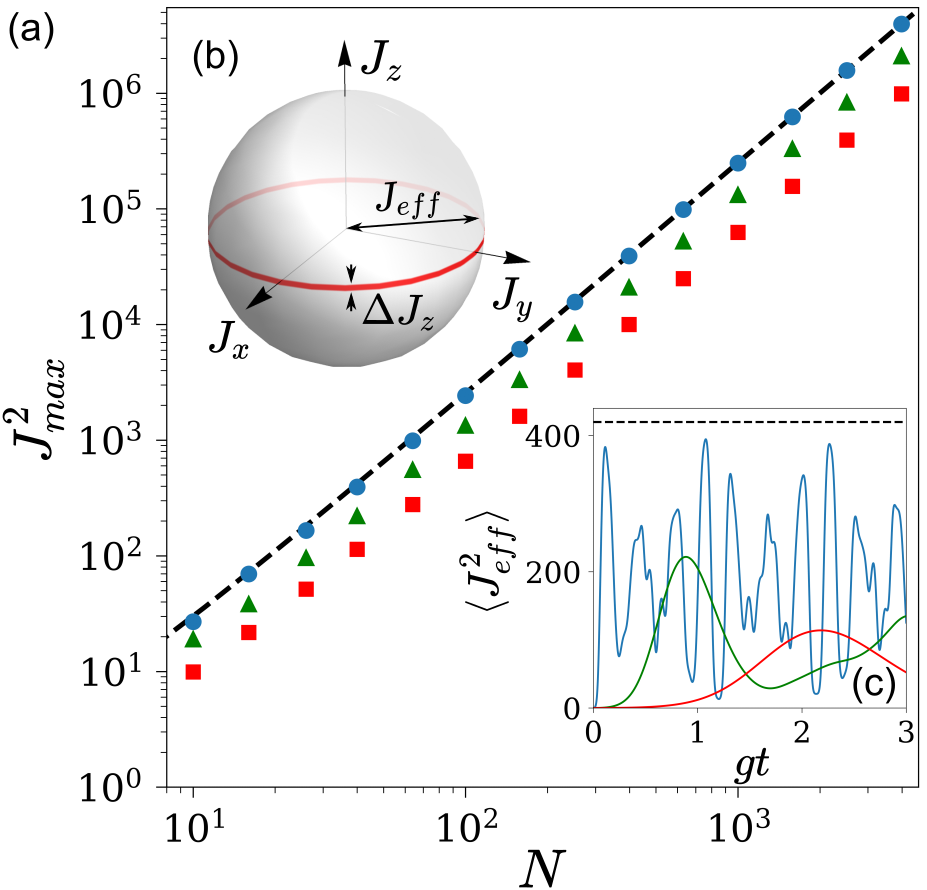}
\caption{Populating the sideband Dicke-like states by unitary evolution via $H_B$ and $H_{ad}$. (a) Maximum $\langle J_{eff}^2\rangle$ reached during temporal evolution, denoted by $J_{max}^2$ (blue dots $H_B$, red squares $H_{ad}$ at $\eta_{eff}=1.6\eta_c$ and green triangles $H_{ad}$ at $\eta_{eff}=3\eta_c$), against the total atom number $N$, with the ideal Dicke state value of $N(N+2)/4$ denoted by the dashed line. (b) The Dicke squeezed state forms a band around the equator of the Bloch sphere, characterized by a large radius $\langle J_{eff}^2\rangle= \langle J_x^2 +J_y^2\rangle$ and a vanishing variance of the ``spin" distribution along the $z$ axis, $\langle (\Delta J_z)^2\rangle$. (c) For unitary evolution with $H_B$ and $H_{ad}$ at $N=40$, $\langle J_{eff}^2\rangle$ (blue line $H_B$, red line $H_{ad}$ at $\eta_{eff}=1.6\eta_c$, green line $H_{ad}$ at $\eta_{eff}=3\eta_c$) performs oscillations in time, nearly reaching the limit value of $N(N+2)/4=420$ (dashed line) for $H_B$. For $H_B$, $g=g_{mod}$, while for $H_{ad}$, $g=\omega_R$. %The minimal $\xi_D$ reached during the temporal evolution, denoted by $\xi_D^{min}$, closely follows the Heisenberg limit of $1/(N+2)$ (dashed line). This indicates that the system very nearly reaches the ideal Dicke state during dynamical evolution.
}\label{Fig:2}
\end{figure}

Dicke squeezed states for the transverse momentum sidebands, depicted on the Bloch sphere in Fig. \ref{Fig:2}b), are characterized by a low $\langle (\Delta J_z)^2\rangle$, large $\langle J_{eff}^2\rangle$ and $\langle J_x\rangle=\langle J_y\rangle=\langle J_z\rangle=0$ \cite{lucke_twin_2011,bucker_twin-atom_2011,zhang_quantum_2014,lucke_detecting_2014,pezze_quantum_2018}. For the two transverse modes, the criterion for Dicke state multiparticle entanglement of identical atoms is given by \cite{toth2009,lucke_detecting_2014}:
\begin{align}\label{eq:xi_gen}
\xi_{gen}^2=(N-1)\frac{\langle (\Delta J_z)^2\rangle}{\langle J_{eff}^2\rangle-N/2}<1,
\end{align}
where $N=N_0+N_++N_-$ is the total number of atoms, taken as constant in our simulations.  

One can easily show that $J_{eff}^2=N_+N_-+(N_++N_-)/2$, such that $\langle J_{eff}^2\rangle=\langle N_+ N_-\rangle+\langle N_++N_-\rangle/2$, where $\langle N_+ N_-\rangle$ is a measure of sideband momentum correlations. For the maximally entangled state, $\langle (\Delta J_z)^2\rangle= 0$ and $\langle  J_{eff}^2\rangle=N(N+2)/4$. In the atomic many-body basis, this maximally multiparticle entangled state is the ideal Dicke state, for even $N$ given by $|0\rangle_0|N/2\rangle_+|N/2\rangle_-$. 

%An alternative measure of Dicke squeezing at large $\langle J_{eff}^2\rangle$, practical in the large $N$ limit since it scales in this ideal case as $1/(N+2)$, is the quantity $\xi_D=N[\langle (\Delta J_z)^2\rangle +0.25]/\langle J_{eff}^2\rangle$ \cite{zhang_quantum_2014}. 

\textit{Continuous translational symmetry of $H_B$ and $H_{cav}$}\label{sec:contsim}. Both $H_B$ and $H_{cav}$ are symmetric to continuous translations along the $x$ axis, which leads to $[\delta N,H_B]=0$ and $[\delta n+\delta N,H_{cav}]=0$, where $\langle J_x\rangle=\langle J_y\rangle=\langle J_z\rangle=0$ for both unitary evolution with $H_B$ and dissipative evolution with $H_{cav}$ \cite{kresic_supplemental_nodate}. The continuous translational symmetry of $H_B$ and $H_{cav}$ is preserved by the density matrix during temporal evolution. These translationally symmetric self-organized states are analogous to maximally amplitude squeezed photonic states, for which the phase of the electric field is undetermined \cite{caves82,walls83}. 

\textit{Unitary evolution under $H_B$ and $H_{ad}$.}\label{sec:results_sc} We now discuss the solutions of the Schrödinger equation for $H_B$ and $H_{ad}$. Due to $[J_z,H_B]=[J_z,H_{ad}]=0$, for a system starting in the state $|N\rangle_0|0\rangle_+|0\rangle_-$, all moments of $J_z$ will be equal to zero at all $t$, leading to $\xi_{gen}^2=0$. %This vanishing of the $J_z$ moments arises from the fact that the transverse sideband states are populated by the atom momentum-conserving process, creating pairs of sideband excitations.

The temporal evolution of $\langle J_{eff}^2\rangle$ is plotted in Fig. \ref{Fig:2}c). The $\langle J_{eff}^2\rangle$ initially rises, and then starts to approximately periodically oscillate in time, with the characteristic timescale $\tau\sim 2\pi/(Ng_{mod})$ for $H_B$ and $\tau\sim 2\pi/(Ng_{cav})=4\pi\eta_c^2/(\omega_R\eta_{eff}^2)$ for $H_{ad}$ \cite{finger23}. For $H_B$, the atoms scatter in and out of the transverse sidebands by absorbing and emitting energy quanta from and to the driving magnetic field, while for $H_{ad}$ one observes sloshing dynamics (see below). The value of $\langle J_{eff}^2\rangle$ oscillates with a large amplitude, nearly reaching 0 at its trough and $N(N+2)/4$ for $H_B$ at its peak. 

The available Hilbert space is for this initial condition significantly reduced, as states keep zero transverse atomic momentum during unitary evolution, allowing for simulations with relatively large $N$. Plotting the highest $J_{eff}^2$ reached during temporal evolution for $H_B$ and $H_{ad}$, denoted by $J_{max}^2$, versus $N$, reveals a rise parallel to the $N(N+2)/4$ line over nearly three orders of magnitude, up to the maximum atom number tractable by the computational resources available, see Fig. \ref{Fig:2}a).

The $N^2$ scaling of $J_{max}^2$, observed for large $N$ during unitary evolution, demonstrates the high efficiency of $H_B$ and $H_{ad}$ in generating momentum entangled atoms, which echoes the efficiency of optical parametric amplifiers in generating photonic entanglement \cite{schnabel17}.

\textit{Dissipative dynamics under $H_{cav}$ and $H_{ad}$.}\label{sec:results_cav}
We first look at the full photon-atom dynamics for both the coherent unitary evolution and for including the cavity photon decay in the Lindblad master equation. Due to conservation of momentum in the combined photon-atom four-wave mixing, we have $[\delta n+\delta N,H_{cav}]= 0$, but $[J_z,H_{cav}]\neq 0$. The Hilbert space is in this problem considerably larger than for $H_B$, which limits the tractable atom number to $N=8$ for the calculations of full quantum dynamics. Parameter values for this case are taken such that maximal Dicke squeezing is observed, and experimentally accessible parameters are discussed in \cite{kresic_supplemental_nodate}.

Due to $[ \delta n+\delta N,H_{cav}]=0$, the variance of $2J_z$ is equal to the variance of $\delta n$ for the unitarily evolving system. This indicates that a reduction of $\xi_{gen}^2$ will benefit from lower $\delta n$ variances, which in general occur when $\langle n_0\rangle$ and $\langle n_\pm\rangle$ are lower, e.g. at larger $|\bar{\Delta}_c|,\:|\bar{\Delta}_c'|$, as long as $\langle J_{eff}^2\rangle$ is large.

\begin{figure}[!t]
\centering
\includegraphics[clip,width=1\columnwidth]{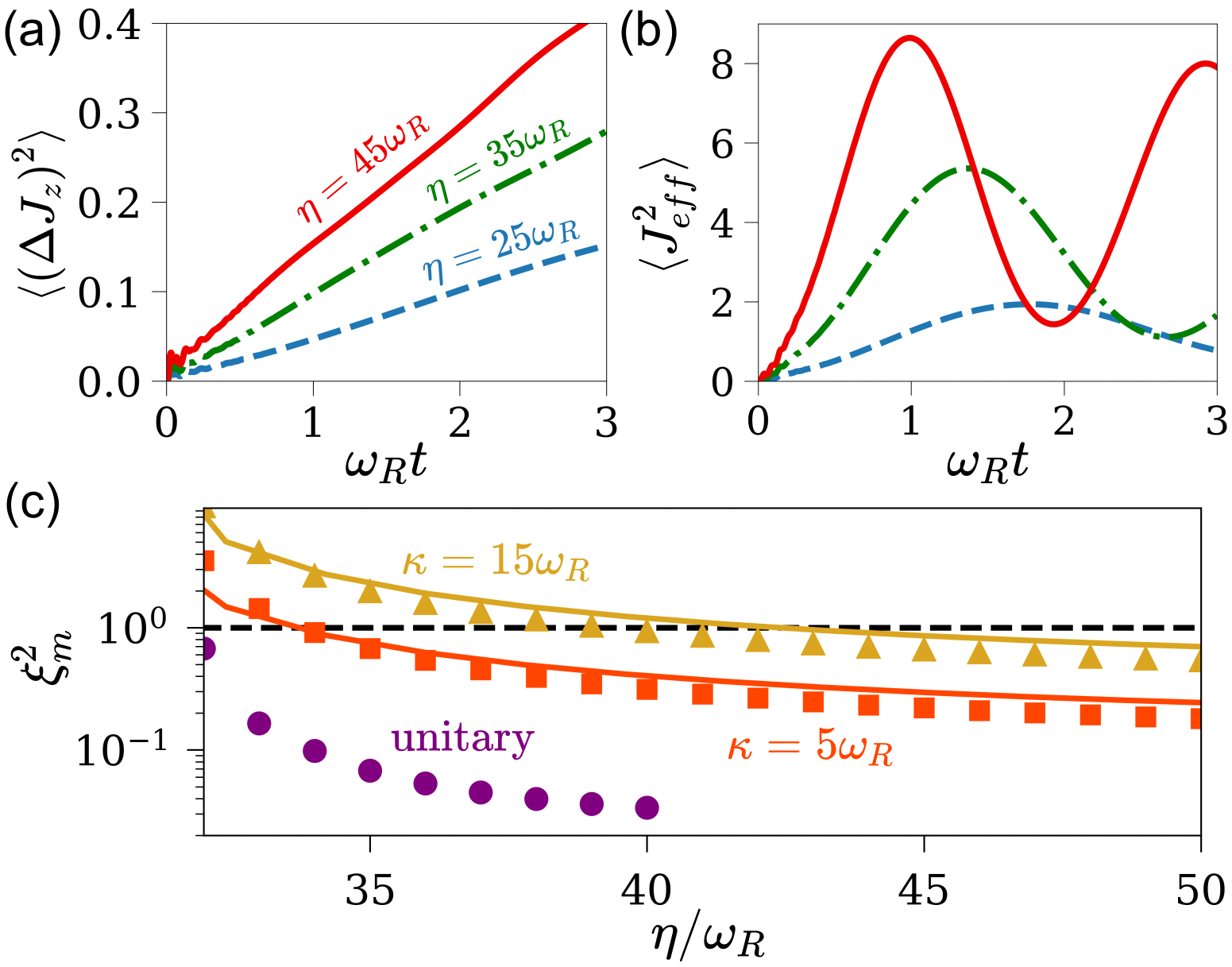}
\caption{Generating multiparticle entanglement in the Dicke-like states via transverse self-organization for the full system Hamiltonian $H_{cav}$. The dissipative dynamical evolution of (a) $\langle (\Delta J_z)^2\rangle$, (b) $\langle J_{eff}^2 \rangle$ for $\eta$ values of $25\omega_R$ (blue, dashed), $35\omega_R$ (green, dot-dashed) and $45\omega_R$ (red, solid). (c) The $\eta$ scan of $\xi_m^2$, given by the lowest $\xi_{gen}^2$ reached during temporal evolution, for the unitary case (purple, dots), and $\kappa=5\omega_R$ (red, squares), $\kappa=15\omega_R$ (yellow, triangles). Crossing of the dashed line ($\xi_{gen}^2=1$) indicates the existence of many particle entanglement \cite{toth2009}. Solid lines in (c) are results for the $H_{ad}$ model with $\eta_{eff}= 0.56\eta$. Simulation parameters: (a), (b) ($\bar{\Delta}_c,\;\bar{\Delta}_c',\; U_0,\;\kappa) = (110,\;-45,\; 10,\;5)\omega_R$, and (c) ($\bar{\Delta}_c,\;\bar{\Delta}_c',\; U_0) = (110,\;-45,\; 10)\omega_R$, with (a)-(c) $N=8$. 
}\label{Fig:3}
\end{figure} 

In Figs.~\ref{Fig:3}a,b) we plot the temporal evolution of the relevant atomic observables for varying the pump strength $\eta$ in the case with dissipation of photons out of the cavity. For the dissipative case, the equality of $\delta N$ and $\delta n$ variances no longer holds and the dissipation of photons out of the cavity makes $\langle (\Delta J_z)^2\rangle$ increase almost linearly with time, with the slope increasing with the pump amplitude $\eta$. %, which indicates that there is still some correlation between $\delta N$ and $\delta n$, since for lower $\eta$ the $\langle n_\pm\rangle$ is lower, leading here to lower $\delta n$ variances. 
This indicates that there is still correlation between $\delta N$ and $\delta n$ even in the dissipative case, i.e.\ the low $\delta n$ variances obtained for lower $\langle n_\pm\rangle$ will lead to lower $\delta N$ variances.

The $\langle J_{eff}^2\rangle$ initially increases and oscillates in time, again indicating sloshing dynamics, with continuous oscillations between the bunched and nearly homogeneous structures in the self-ordered atomic lattice (see also \cite{nagy_dicke-model_2010,tesio_self-organization_2014}). Increasing the pump $\eta$, both the growth rate and maximum value of $\langle J_{eff}^2\rangle$ are increased. The growth rate of $\langle J_{eff}^2\rangle$ increases faster with distance to threshold than the growth rate of $\langle (\Delta J_z)^2\rangle$, which, from Eq.~(\ref{eq:xi_gen}), leads to a decrease in $\xi_{gen}^2$ for larger $\eta$. Increasing the $\eta$ values several times above the semiclassically calculated threshold $\eta_c=[-\omega_R(\bar{\Delta}_c^2+\kappa^2)(\bar{\Delta}_c'^2+\kappa^2)/(4NU_0^2\bar{\Delta}_c')]^{1/2}\approx 13\omega_R$ \cite{kresic_supplemental_nodate}, the $\xi_{m}^2$, given by the lowest $\xi_{gen}^2$ attained during temporal evolution, reaches values of $\xi_{m}^2=0.03$ for the unitary and $\xi_{m}^2=0.18$ for the dissipative case, see Fig. \ref{Fig:3}c).  %The lowest $\xi_{m}^2$ values for these results are $\xi_{m}^2=0.03$ for the unitary and $\xi_{m}^2=0.18$ for the dissipative case. The maximal $\eta$ available for these simulations is limited by the size of the numerically tractable Hilbert space. 
We have also performed simulations of dissipative dynamics with adiabatically eliminated photonic modes, described by a Lindblad master equation with $H_{ad}$ and decay rate $\gamma=U_0^2\eta_{eff}^2 \kappa/[(\bar{\Delta}_c'^2+\kappa^2)(\bar{\Delta}_c^2+\kappa^2)]$. The $H_{ad}$ calculations with $\eta_{eff}=0.56\eta$ reproduce well the squeezing for the full model, as shown in Fig. \ref{Fig:3}c). The rescaling of $\eta_{eff}$ with respect to $\eta$ is a consequence of neglecting the nonlinear saturation and on-axis mode depletion terms in deriving $H_{ad}$ \cite{kresic_supplemental_nodate}.

\begin{figure}[!t]
\centering
\includegraphics[clip,width=1\columnwidth]{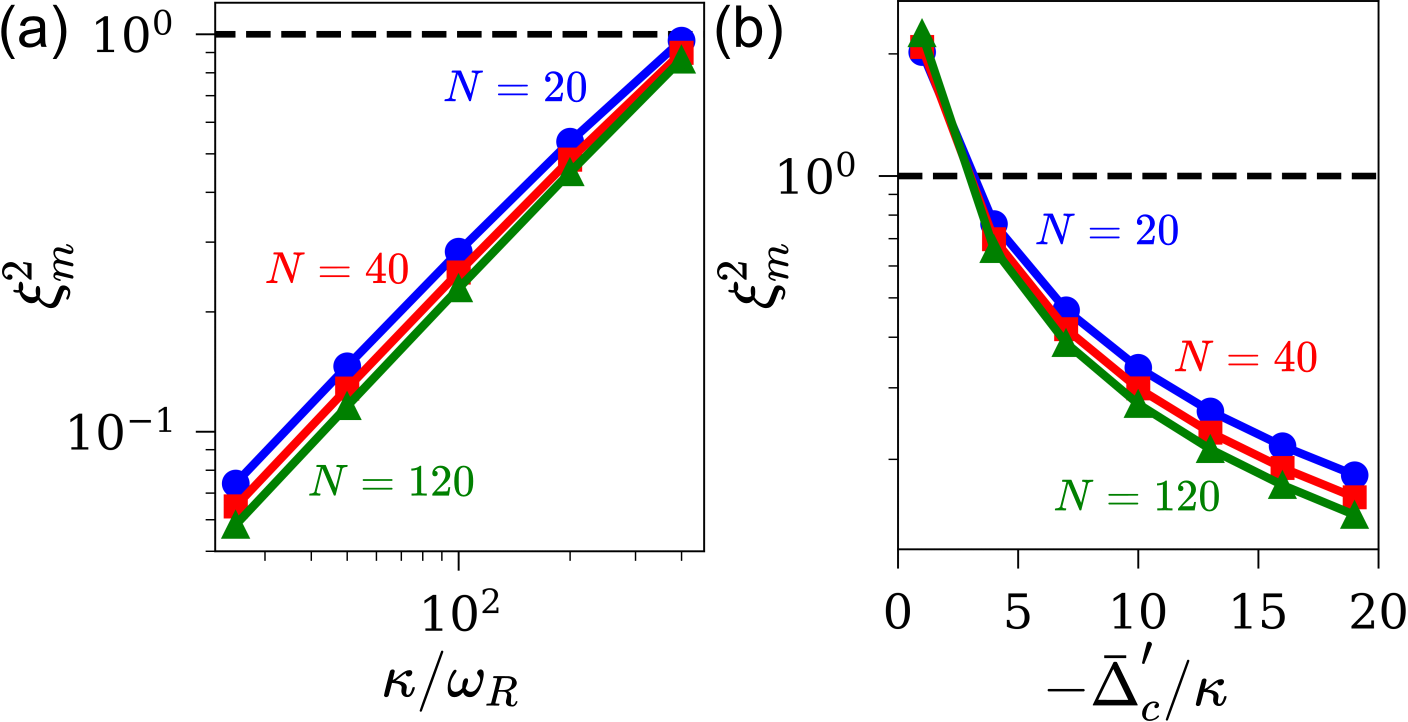}
\caption{Entanglement generation for a dissipative cavity with adiabatically eliminated photonic fields, described by the Lindblad master equation with $H_{ad}$ and decay rate $\gamma$ (see text). (a) Cavity decay rate $\kappa$ scan of $\xi_m^2$ (see text) for fixed $(\bar{\Delta}_c,\:\bar{\Delta}_c')=(1400,\:-1200)\omega_R$, and $N=20$ (blue dots), $N=40$ (red squares) and $N=120$ (green triangles). (b) Sideband detuning $\bar{\Delta}_c'$ scan of $\xi_m^2$ for $N$ values as in (a), and fixed $\kappa=3\times 10^4\omega_R$, with $\bar{\Delta}_c=|\bar{\Delta}_c'|+4\kappa$ for each point. The dashed line indicates $\xi_{gen}^2=1$. Solid lines are guide to the eyes. Simulation parameters: $U_0=10^{-3}\omega_R$, $\eta_{eff}=1.6\eta_c$. 
}\label{Fig:4}
\end{figure}

For higher $N$ calculations, reachable with the $H_{ad}$ model, the $\xi_m^2$ again decreases for decreasing the decay rate $\kappa$ at fixed detunings $\bar{\Delta}_c,\:\bar{\Delta}_c'$ and $\eta_{eff}/\eta_c$ [see Fig. \ref{Fig:4}a)], which is a consequence of increasing the ratio $-g_{cav}/\gamma=2|\bar{\Delta}_c'|/\kappa$, leading towards coherent cavity dynamics similarly to the single mode cavity case \cite{davis19,finger23}. The parameters $\bar{\Delta}_c'/\kappa$ and $\eta_{eff}/\eta_c$ completely determine the open system dynamics for $H_{ad}$ with $N$ atoms. Intriguingly, significant squeezing can be achieved in this model for large $-\bar{\Delta}_c'/\kappa$ even with ``bad cavity" parameters. Indeed, using the finesse $F=250$ and a free spectral range of FSR $= 1.5$ GHz, with $\kappa/(2\pi)=\mbox{FSR}/(2F)= 3$ MHz and $\omega_R=2\pi\times 0.1$ kHz, we demonstrate in Fig. \ref{Fig:4}b) that nearly $10$ dB squeezing is attainable for $N=120$ atoms. 

%\textbf{The influence of cavity dissipation on transient dynamics becomes negligible as $\bar{\Delta}_c'/\kappa$ increases. The $N^2$ scaling of $J_{max}^2$, demonstrated in Fig. \ref{Fig:2}a) for unitary evolution with $H_{ad}$, will therefore lead to Heisenberg scaling for large $\bar{\Delta}_c'/\kappa$. }

%The $\xi_m^2$ also decreases for increasing the number of atoms $N$. Noting that the largest $\omega_R$ for the $^{87}$Rb D2-line is $2\pi\times 3.77$ kHz, and state of the art cavities can achieve $\kappa=2\pi\times (4.5-10^3)$ kHz \cite{baumann_dicke_2010,wolke12,kollar_supermode-density-wave-polariton_2017}, significant entanglement generation should be well within experimental reach.}

%\textit{Discussion.} 
\textit{Conclusion.}\label{sec:conclusion} Momentum entanglement of ultracold atoms holds great potential for enhancing matter wave interferometry \cite{szigeti21}. We have theoretically and numerically explored two methods for producing momentum multiparticle entangled (Dicke squeezed) states of atoms with one internal ground state (spin-0), by nonlinear self-organization in driven systems. For $H_B$ the drive is an oscillating magnetic field, and the method is based on the experiment of Ref.~\cite{zhang_pattern_2020}. A novel model $H_{cav}$, based on laser driving of a two-level optical transition of a BEC with ring cavity feedback, is also developed and explored. Although entanglement occurs here in the transient dynamical regime, it can readily be shown that switching off the driving field at the moment of optimal squeezing preserves the momentum distribution and thus also the entanglement for a longer time \cite{kresic_supplemental_nodate}. The three mode picture, for which the higher order momentum sideband population is considered negligible, can prove incomplete when scattering to higher order modes is strong, which is explored in \cite{kresic_supplemental_nodate}.

Experimental determination of $\langle (\Delta J_z)^2\rangle$ and $\langle J_{eff}^2\rangle$ relies on assessing the momentum correlations via $\langle N_+ N_-\rangle$. Absorption imaging of the atomic momentum distribution with a highly efficient camera can be used to measure $N_+ N_-$ for each experimental realization, and $\langle N_+ N_-\rangle$ can then be determined by taking the average for many realizations. Alternative ways of measuring atomic momentum correlations are given e.g. in \cite{bucker_twin-atom_2011,borselli21}. Note that Dicke squeezing generation described in this Letter is based on continuous symmetry of $H_B$ and $H_{cav}$, which is also present for some prominent recent models describing self-organization of ultracold atoms~\cite{leonard_supersolid_2017,schuster2020}. %It should thus be possible to observe multiparticle entanglement generation via self-organization in experimental setups of Refs.~\cite{leonard_supersolid_2017,schuster2020}, along with the setup of Ref.~\cite{zhang_pattern_2020}.} 

Moreover, since the calculations show significant momentum squeezing is available even for low finesse values, a relatively simple setup with a nearly degenerate cavity external to the vacuum chamber may be sufficient to realize it experimentally. Another highly intriguing prospect is the possibility of generating correlations in thermal and ultracold atomic degrees of freedom in a single mirror feedback setup \cite{dalessandro91,labeyrie_optomechanical_2014,kresic_spontaneous_2018,labeyrie_18,kresic_inversion-symmetry_2019,ackemann_self-organization_2021,labeyrie2022} or using counterpropagating beams \cite{yariv77,greenberg_bunching-induced_2011,firth17}.

\textit{Acknowledgements.} We thank Tobias Donner, Oliver Diekmann, Helmut Ritsch, Berislav Bu\v ca, Paul Griffin, and Onur Hosten for helpful discussions. The work of I. K. was funded by the Austrian Science
Fund (FWF) Lise Meitner Postdoctoral Fellowship M3011 and an ESQ Discovery grant from the Austrian Academy of Sciences (ÖAW). The dynamical evolution equations were solved numerically by using the open-source framework QuantumOptics.jl \cite{kramer_quantumopticsjl_2018}. The computational results presented here have been achieved using the Vienna Scientific Cluster (VSC).

\bibliography{references}
\clearpage
\newpage

\end{document}

%\begin{document}

\title{Supplemental Material - Generating multiparticle entangled states by self-organization of driven ultracold atoms}

\author{Ivor Kre\v{s}i\'{c}} \email{ivor.kresic@tuwien.ac.at} 
\affiliation{Institute for Theoretical Physics, Vienna University of Technology (TU Wien), Vienna, A–1040, Austria}
\affiliation{Centre for Advanced Laser Techniques, Institute of Physics, Bijeni\v{c}ka cesta 46, 10000, Zagreb, Croatia}
\author{Gordon R. M. Robb} 
\affiliation{SUPA and Department of Physics, University of Strathclyde, Glasgow G4 0NG, Scotland, UK}
\author{Gian-Luca Oppo} 
\affiliation{SUPA and Department of Physics, University of Strathclyde, Glasgow G4 0NG, Scotland, UK}
\author{Thorsten Ackemann} 
\affiliation{SUPA and Department of Physics, University of Strathclyde, Glasgow G4 0NG, Scotland, UK}

%\date{\today}
\maketitle

\section{Derivation of $H_{B}$ and $H_{cav}$}
The derivation of $H_B$ is given for 2D clouds in the Methods section of Ref. \cite{zhang2020}, and can be readily simplified to the 1D case. The starting Hamiltonian is for this case given by:
\begin{align}\label{eq:hamb_1}
h_B=\int_V d^3r\psi^\dagger(\mathbf{r})\left[\frac{\mathbf{p}^2}{2m}+ \frac{g(t)}{2}\psi^\dagger(\mathbf{r}) \psi(\mathbf{r})\right]\psi(\mathbf{r}),
\end{align}
with $g(t)=\frac{4\pi\hbar^2}{m} [a_{dc}+a_{mod}\sin(\omega_{mod}t)]$, where $a_{dc}$ is a small offset scattering length required for the stability of the condensate. In the experiment of Ref \cite{zhang2020}, $a_{mod}=22.5 a_{dc}$, and thus $g(t)$ is approximated as $g(t)\approx \frac{4\pi\hbar^2}{m}a_{mod}\sin(\omega_{mod}t)$. Neglecting here the higher order momentum terms (see section ``Suppression of higher order sideband excitation"), the atomic field operator is approximated as:
\begin{align}\label{eq:atomfield2}
\psi(\mathbf{r})=\frac{1}{\sqrt{V}}\left(b_0+b_+e^{ik_fx}+b_-e^{-ik_fx}\right),
\end{align}
where $b_j$ is the bosonic annihilation operator of the $j$-th transverse atomic momentum mode and $k_f$ is the critical wavenumber. The Hamiltonian now has the form:
\begin{align}\label{eq:hamb_2}
h_B=\varepsilon(N_++N_-)+\frac{i\pi\hbar^2a_{mod}}{m}\int_V d^3r\psi^\dagger(\mathbf{r})\psi^\dagger(\mathbf{r}) \left( e^{-i\omega_{mod}t} -e^{i\omega_{mod}t}\right)\psi(\mathbf{r})\psi(\mathbf{r}),
\end{align}
where $\varepsilon_\pm=\hbar^2k_f^2/2m=\varepsilon$. We now use $U_{rot}(t)=\exp[i\varepsilon (N_++N_-)t/\hbar]$ to transform into the interaction picture where the unitary evolution for the state $|\psi(t)\rangle_I=U_{rot}(t)|\psi(t)\rangle_S$ is given by:
\begin{align}\label{eq:eqhamb1}
i\hbar \partial_t |\psi(t)\rangle_I=i\hbar \partial_t U_{rot}(t)|\psi(t)\rangle_S=\left[-\varepsilon(N_++N_-)+U_{rot}(t)h_B U_{rot}^{-1}(t)\right]|\psi(t)\rangle_I.
\end{align}
From the above equation we arrive to the interaction Hamiltonian
\begin{align}\label{eq:eqhamb2}
H_B=U_{rot}(t)\left[\frac{i\pi\hbar^2a_{mod}}{m}\int_V d^3r\psi^\dagger(\mathbf{r})\psi^\dagger(\mathbf{r}) \left( e^{-i\omega_{mod}t} -e^{i\omega_{mod}t}\right)\psi(\mathbf{r})\psi(\mathbf{r})\right] U_{rot}^{-1}(t).
\end{align}
One can now use the relations $U_{rot}(t)b_0 U_{rot}^{-1}(t)=b_0,\:U_{rot}(t)b_\pm U_{rot}^{-1}(t)=e^{-i\varepsilon t/\hbar}b_\pm$, where similar relations are derived also for the translation operator in section ``Continuous translational symmetry of $H_B$ and $H_{cav}$" below. At the critical modulation frequency $\omega_{mod}=2\varepsilon/\hbar=\hbar k_f^2/m$, it can readily be shown that for neglecting the fast-oscillating terms in Eq. (\ref{eq:eqhamb2}), one arrives at the interaction Hamiltonian in the rotating-wave approximation: 
\begin{align}\label{eq:eqhambfinal}
H_B=i\hbar g_{mod} b_+^\dagger b_-^\dagger b_0 b_0+\mbox{H.c.},
\end{align}
where $g_{mod}=2\pi\hbar a_{mod}/mV$. Note that taking $g(t)\approx \frac{4\pi\hbar^2}{m}a_{mod}\cos(\omega_{mod}t)$, leads to
\begin{align}\label{eq:eqhamb2prime}
H_B'=U_{rot}(t)\left[\frac{\pi\hbar^2a_{mod}}{m}\int_V d^3r\psi^\dagger(\mathbf{r})\psi^\dagger(\mathbf{r}) \left( e^{-i\omega_{mod}t} +e^{i\omega_{mod}t}\right)\psi(\mathbf{r})\psi(\mathbf{r})\right] U_{rot}^{-1}(t),
\end{align}
which results in:
\begin{align}\label{eq:eqhambfinalprime}
H_B'=\hbar g_{mod} b_+^\dagger b_-^\dagger b_0 b_0+\mbox{H.c.}.
\end{align}

We now provide the details of the derivation of $H_{cav}$. A cigar-shaped zero-temperature Bose-Einstein condensate (BEC) is placed in an effectively plano-planar ring cavity of effective diffractive length $L$ with one lossy and three perfectly reflecting mirrors ($\kappa$ - cavity photon decay rate), which is pumped by a coherent electric field with pump strength $\eta$ at frequency $\omega$ (see Fig. \ref{Fig:meanfield}a) and section ``Experimental design of the ring cavity setup" for details). A strong confinement along the $y$ and $z$ axes allows to restrict the analysis to 1D structures. The pump drive excites on-axis running waves with spatial profile $e^{ik_0z}$, with the spontaneously generated sidebands having the profile $e^{ik_0z}e^{\pm iq_cx}$, where $k_0=2\pi/\lambda_0$ is the cavity longitudinal mode (also called the on-axis or zero-order mode) wavenumber, $q_c=2\pi/\Lambda_c$ and $\Lambda_c$ is the pattern lengthscale, tunable via Fourier filtering of intracavity light \cite{jensen98}.

Following Refs. \cite{lugiato_quantum_1992,grynberg_quantum_1993,gatti_spatial_2001}, we take the electric field modes as:
\begin{align}\label{eq:elfield}
E(\mathbf{r})=a_0e^{ik_0z}+a_+e^{ik_0z}e^{iq_cx}+a_-e^{ik_0z}e^{-iq_cx},
\end{align}
where $a_j$ are the photonic annihilation operators in the $j$-th mode. The atomic field operator is given by:
\begin{align}\label{eq:atomfield}
\psi(\mathbf{r})=\frac{1}{\sqrt{V}}\left(b_0+b_+e^{iq_cx}+b_-e^{-iq_cx}\right),
\end{align}
where $b_j$ is the bosonic annihilation operator of the $j$-th transverse atomic momentum mode. The effective many-body Hamiltonian for the photons and atomic motional states can be derived from the Jaynes-Cummings model (see e.g. \cite{maschler_ultracold_2008,tesio_theory_2014,torggler_adaptive_2014}). For a two-level optical transition, using the dipole and rotating wave approximations in the low saturation (far-detuned) limit, we therefore have:
\begin{equation}\label{eq:ham1}
\begin{aligned}
% H =-\hbar( \Delta_ca_0^\dagger a_0+\Delta_c'\sum_{j=\pm} a_j^\dagger a_j)+i\hbar(\eta a_0^\dagger-\eta^* a_0)  \\
H_{cav} =& -\hbar \Delta_cn_0-\hbar\Delta_c'(n_++n_-)+i\hbar(\eta a_0^\dagger-\eta^* a_0)+ \int_V d^3r\psi^\dagger(\mathbf{r})\left[\frac{\mathbf{p}^2}{2m}+ \hbar U_0E^\dagger(\mathbf{r}) E(\mathbf{r})\right]\psi(\mathbf{r}),
\end{aligned}
\end{equation}
where $\Delta_c=\omega-\omega_0,\;\Delta_c'=\omega-\omega_0'$ are the pump detunings from the on-axis and sideband cavity modes, respectively,
%(we put $\Delta_c'=\Delta_c$), 
$n_{0}=a_{0}^\dagger a_{0},\: n_{\pm}=a_{\pm}^\dagger a_{\pm}$,
%$\Delta_c'=\omega-\omega_0'$ are the laser-cavity detunings for the on-axis and sideband modes, respectively, 
$U_0=g_0^2/\Delta_a$ is the single atom light shift, $\Delta_a=\omega-\omega_a$ is the laser-atom detuning, and $g_0$ is the atom-cavity coupling strength. In writing Eq. (\ref{eq:ham1}) we have neglected the random collisions between the atoms in the dilute BEC cloud, as we are here interested on highlighting the consequences of light-matter interaction. The number of atoms is fixed and given by $N=N_0+N_++N_-$, where $N_{0}=b_{0}^\dagger b_{0},\:N_{\pm}=b_{\pm}^\dagger b_{\pm}$.

We insert Eqs. (\ref{eq:elfield}) and (\ref{eq:atomfield}) into Eq. (\ref{eq:ham1}) for real-valued pump rate $\eta$ and perform the integration over the BEC cloud volume $V$ to get the effective total Hamiltonian $H_{cav} = H_0+H_{FWM}^{(1)}+H_{FWM}^{(2)}$,
where:
\begin{equation}\label{eq:h0}
\begin{aligned}
%\begin{split}
H_0 &= -\hbar\bar{\Delta}_cn_0-\hbar\bar{\Delta}_c'(n_++n_-)
+\hbar\omega_R(N_++N_-)+i\hbar(\eta a_0^\dagger- \eta^* a_0),
%\end{split}
\end{aligned}
\end{equation}
and the four-wave mixing terms are:
\begin{align}\label{eq:fwmterms_1}
\begin{split}
H_{FWM}^{(1)}& = \hbar U_0 [(a_+^\dagger b_-^\dagger + a_-^\dagger b_+^\dagger)a_0 b_0 +\mbox{H.c.}] 
+\hbar U_0 [a_0^\dagger(b_+^\dagger a_+ +b_-^\dagger a_-)b_0 +\mbox{H.c.}], 
\end{split} \\
H_{FWM}^{(2)}& = \hbar U_0(a_+^\dagger a_- b_-^\dagger
b_++\mbox{H.c.})\label{eq:fwmterms_2},
\end{align}
where $\bar{\Delta}_c=\Delta_c-NU_0$, $\bar{\Delta}_c'=\Delta_c'-NU_0$ and $\hbar\omega_{R}=(\hbar q_c)^2/2m$ is the transverse recoil energy.

Generation of transverse sidebands via $H_{FWM}^{(1)}$ can be explained by the momentum conserving processes illustrated in Fig. 1d) of the main text. The $H_{FWM}^{(2)}$ describes the secondary wave mixing process for stripe patterns, in which a scattering of a photon sideband with $\pm q_c$ into the mode with $\mp q_c$ leads to a transition of an atom from the state with transverse momentum $\mp \hbar q_c$ into the state with transverse momentum $\pm \hbar q_c$. This process leads to saturation of the sideband mode population far above threshold \cite{ackemann2021}.

\section{Mean field evolution equations for $H_{cav}$}
We first look at the temporal evolution of the field operators \cite{breuer02}. The Heisenberg equation for an operator $O(t)$ has the form:
\begin{align}\label{eq:heis1}
\frac{dO}{dt}=\frac{i}{ \hbar }[H,O].
\end{align}
The nonunitary dynamics, including the dissipation of photons from the cavity, can be described by the Lindblad-type evolution for the expectation value $\langle O\rangle=\mbox{Tr}(O\rho (t))$, via the equation \cite{albert14}:
\begin{align}\label{eq:lind1}
\frac{d\langle O\rangle }{dt}=\langle \mathcal{L}^\dagger[O]\rangle=\frac{i}{\hbar}\langle [H_{cav},O]\rangle +\kappa\sum_{j=0,\pm}\langle 2a_j^\dagger O a_j-a_j^\dagger a_j O-O a_j^\dagger a_j\rangle.
\end{align}
We here use the ring cavity Hamiltonian $H_{cav}=H_0+H_{FWM}^{(1)}+H_{FWM}^{(2)}$, with the three parts given by:
\begin{align}\label{eq:hamiltonian_again}
 & H_0  =  -\hbar\bar{\Delta}_cn_0-\hbar\bar{\Delta}_c'(n_++n_-)+\hbar\omega_R(N_++N_-)+i\hbar ( \eta a_0^\dagger- \eta^* a_0),\\
& H_{FWM}^{(1)} =  \hbar U_0 [(a_+^\dagger b_-^\dagger + a_-^\dagger b_+^\dagger)a_0 b_0 +a_0^\dagger(b_+^\dagger a_+ +b_-^\dagger a_-)b_0 +\mbox{H.c.}], \\
& H_{FWM}^{(2)} =  \hbar U_0(a_+^\dagger a_- b_-^\dagger b_++\mbox{H.c.}),
\end{align}
and the commutation relations of the bosonic modes for photons and atoms: $[a_j,a_k^\dagger]=\delta_{j,k}$, $[a_j^\dagger,a_k^\dagger]=[a_j,a_k]=0$, and  $[b_j,b_k^\dagger]=\delta_{j,k}$, $[b_j^\dagger,b_k^\dagger]=[b_j,b_k]=0$, respectively, where $j,k=0,+,-$. For the photonic modes, the Eq. (\ref{eq:lind1}) now gives:
\begin{align}\label{eq:photlang}
\langle \dot{a}_0\rangle &=(i\bar{\Delta}_c-\kappa)\langle a_0\rangle -iU_0\langle(b_+^\dagger a_++b_-^\dagger a_-)b_0+b_0^\dagger(a_+b_-+a_-b_+) \rangle+\eta,\\
\langle\dot{a}_+\rangle &=(i\bar{\Delta}_c'-\kappa)\langle a_+\rangle -iU_0\langle(b_-^\dagger b_0+b_0^\dagger b_+)a_0+a_-b_-^\dagger b_+\rangle,\\
\langle\dot{a}_-\rangle &=(i\bar{\Delta}_c'-\kappa)\langle a_-\rangle -iU_0\langle(b_+^\dagger b_0+b_0^\dagger b_-)a_0+a_+b_+^\dagger b_-\rangle,
\end{align}
while for the atomic momentum modes, the Eq. (\ref{eq:heis1}) gives:
\begin{align}\label{eq:atomheis}
\langle\dot{b}_0\rangle &=-iU_0\langle a_0^\dagger(a_+b_-+a_-b_+)+(a_+^\dagger b_++a_-^\dagger b_-)a_0)\rangle,\\
\langle\dot{b}_+\rangle &=-i\omega_R \langle b_+\rangle -iU_0\langle(a_-^\dagger a_0+a_0^\dagger a_+)b_0+a_-^\dagger a_+ b_-\rangle,\\
\langle\dot{b}_-\rangle &=-i\omega_R \langle b_-\rangle -iU_0\langle(a_+^\dagger a_0+a_0^\dagger a_-)b_0+a_+^\dagger a_- b_+\rangle.
\end{align}
Taking now the expectation values of the right- and left-hand sides, writing $\langle O_1O_2O_3\rangle\to \langle O_1\rangle \langle O_2\rangle \langle O_3\rangle$, and substituting $\langle a_j\rangle\to \sqrt{N}\alpha_j(t),\:\langle b_j\rangle\to \sqrt{N}\beta_j(t)$, we get the mean field dynamical equations:
\begin{align}
\dot{\alpha}_0 &=(i\bar{\Delta}_c-\kappa)\alpha_0-iu_0[(\beta_+^* \alpha_++\beta_-^* \alpha_-)\beta_0+\beta_0^*(\alpha_+\beta_-+\alpha_-\beta_+) ]+y,\label{eq:photlang_mean}\\
\dot{\alpha}_+ &=(i\bar{\Delta}_c'-\kappa)\alpha_+-iu_0[(\beta_-^* \beta_0+\beta_0^* \beta_+)\alpha_0+\alpha_-\beta_-^* \beta_+],\\
\dot{\alpha}_- &=(i\bar{\Delta}_c'-\kappa)\alpha_--iu_0[(\beta_+^* \beta_0+\beta_0^* \beta_-)\alpha_0+\alpha_+\beta_+^* \beta_-],
\end{align}
where $u_0=NU_0$, $y=\eta/\sqrt{N}$, and
\begin{align}
\dot{\beta}_0 &=-iu_0[\alpha_0^*(\alpha_+\beta_-+\alpha_-\beta_+)+(\alpha_+^* \beta_++\alpha_-^* \beta_-)\alpha_0)],\\
\dot{\beta}_+ &=-i\omega_R \beta_+-iu_0[(\alpha_-^* \alpha_0+\alpha_0^* \alpha_+)\beta_0+\alpha_-^* \alpha_+ \beta_-],\\
\dot{\beta}_- &=-i\omega_R \beta_--iu_0[(\alpha_+^* \alpha_0+\alpha_0^* \alpha_-)\beta_0+\alpha_+^* \alpha_- \beta_+]\label{eq:atomheis_mean}.
\end{align}

\section{Mean field threshold for $H_{cav}$}
We now look at the steady state limit of the above mean field dynamical equations. Writing now $\alpha_j(t)\to \alpha_j$ and $\beta_j(t)\to \beta_j$, we get the equations: 
\begin{align}\label{eq:photlang_meanst}
0 &=(i\bar{\Delta}_c-\kappa)\alpha_0-iu_0[(\beta_+^* \alpha_++\beta_-^* \alpha_-)\beta_0+\beta_0^*(\alpha_+\beta_-+\alpha_-\beta_+) ]+y,\\
0 &=(i\bar{\Delta}_c'-\kappa)\alpha_+-iu_0[(\beta_-^* \beta_0+\beta_0^* \beta_+)\alpha_0+\alpha_-\beta_-^* \beta_+],\\
0 &=(i\bar{\Delta}_c'-\kappa)\alpha_--iu_0[(\beta_+^* \beta_0+\beta_0^* \beta_-)\alpha_0+\alpha_+\beta_+^* \beta_-],
\end{align}
and
\begin{align}\label{eq:atomheis_meanst}
0 &=u_0[\alpha_0^*(\alpha_+\beta_-+\alpha_-\beta_+)+(\alpha_+^* \beta_++\alpha_-^* \beta_-)\alpha_0)],\\
0 &=\omega_R \beta_++u_0[(\alpha_-^* \alpha_0+\alpha_0^* \alpha_+)\beta_0+\alpha_-^* \alpha_+ \beta_-],\\
0 &=\omega_R \beta_-+u_0[(\alpha_+^* \alpha_0+\alpha_0^* \alpha_-)\beta_0+\alpha_+^* \alpha_- \beta_+].
\end{align}
At threshold, we neglect all terms square or higher order in the sidebands. We choose a real-valued $\beta_0=\sqrt{1-|\beta_+|^2-|\beta_-|^2}\approx 1$ and find the homogeneous field amplitude of the on-axis mode to be
\begin{align}\label{eq:alpha0}
\alpha_0=\frac{y}{\sqrt{\bar{\Delta}_c^2+\kappa^2}}e^{i\arctan(\bar{\Delta}_c/\kappa)}.
\end{align}
One can then easily calculate that in this approximation:
\begin{align}\label{eq:betapm}
\beta_\pm = -\frac{u_0}{\omega_R}(\alpha_\pm^*\alpha_0+\alpha_\mp\alpha_0^*).
\end{align}
Inserting this relation in the equations for the fields, we get:
\begin{align}\label{eq:alphapm}
0 &=(i\bar{\Delta}_c'-\kappa)\alpha_++2i\frac{u_0^2}{\omega_R}(\alpha_+\alpha_0^*+\alpha_-^*\alpha_0)\alpha_0,\\
0 &=(i\bar{\Delta}_c'-\kappa)\alpha_-+2i\frac{u_0^2}{\omega_R}(\alpha_+^*\alpha_0+\alpha_-\alpha_0^*)\alpha_0.
\end{align}
By inserting the complex conjugate of
\begin{align}\label{eq:eqforalphapm}
\alpha_+\alpha_0^*+\alpha_-^*\alpha_0=-\frac{(i\bar{\Delta}_c'-\kappa)}{2i\frac{u_0^2}{\omega_R}}\frac{\alpha_+}{\alpha_0}
\end{align}
into the second equation of the above, we get:
\begin{align}\label{eq:eqforalpham}
\alpha_-=\frac{i\bar{\Delta}_c'+\kappa}{i\bar{\Delta}_c'-\kappa}\frac{\alpha_0\alpha_+^*}{\alpha_0^*}.
\end{align}
As the sidebands have equal amplitudes, we can write:
\begin{align}\label{eq:sidebands}
\alpha_\pm=Ae^{i\chi_\pm},\: \beta_\pm=Be^{i\epsilon_\pm }.
\end{align}
From (\ref{eq:betapm}), we then get that at threshold $\bar{\epsilon}=\epsilon_++\epsilon_-=0$, while Eq. (\ref{eq:eqforalpham}) gives $\bar{\chi}=\chi_++\chi_-=2[\arctan(\bar{\Delta}_c/\kappa)+\arctan(\bar{\Delta}_c'/\kappa)+\arg y]$. Inserting the complex conjugate of Eq. (\ref{eq:eqforalpham}) into the Eq. (\ref{eq:alphapm}), we get for the critical intracavity field:
\begin{align}\label{eq:critalpha0}
|\alpha_0^c|^2=-\frac{\omega_R(\bar{\Delta}_c'^2+\kappa^2)}{4u_0^2\bar{\Delta}_c'}
\end{align}
which gives the threshold for the real-valued critical input electric field amplitude $y_c$:
\begin{align}\label{eq:crity}
y_c^2=-\frac{\omega_R(\bar{\Delta}_c^2+\kappa^2)(\bar{\Delta}_c'^2+\kappa^2)}{4u_0^2\bar{\Delta}_c'},
\end{align}
which leads to 
\begin{align}\label{eq:criteta}
\eta_c=\sqrt{\frac{-\omega_R(\bar{\Delta}_c^2+\kappa^2)(\bar{\Delta}_c'^2+\kappa^2)}{4NU_0^2\bar{\Delta}_c'}}.
\end{align}
The structure of the expression (\ref{eq:criteta}) for $\eta_c$ bears similarity to the threshold for the transversely pumped single mode cavity, given by $\sqrt{-\omega_R(\bar{\Delta}_c^2+\kappa^2)/(2N\bar{\Delta}_c)}$ \cite{nagy2011}. 

\begin{figure}[!t]
\centering
\includegraphics[clip,width=0.85\columnwidth]{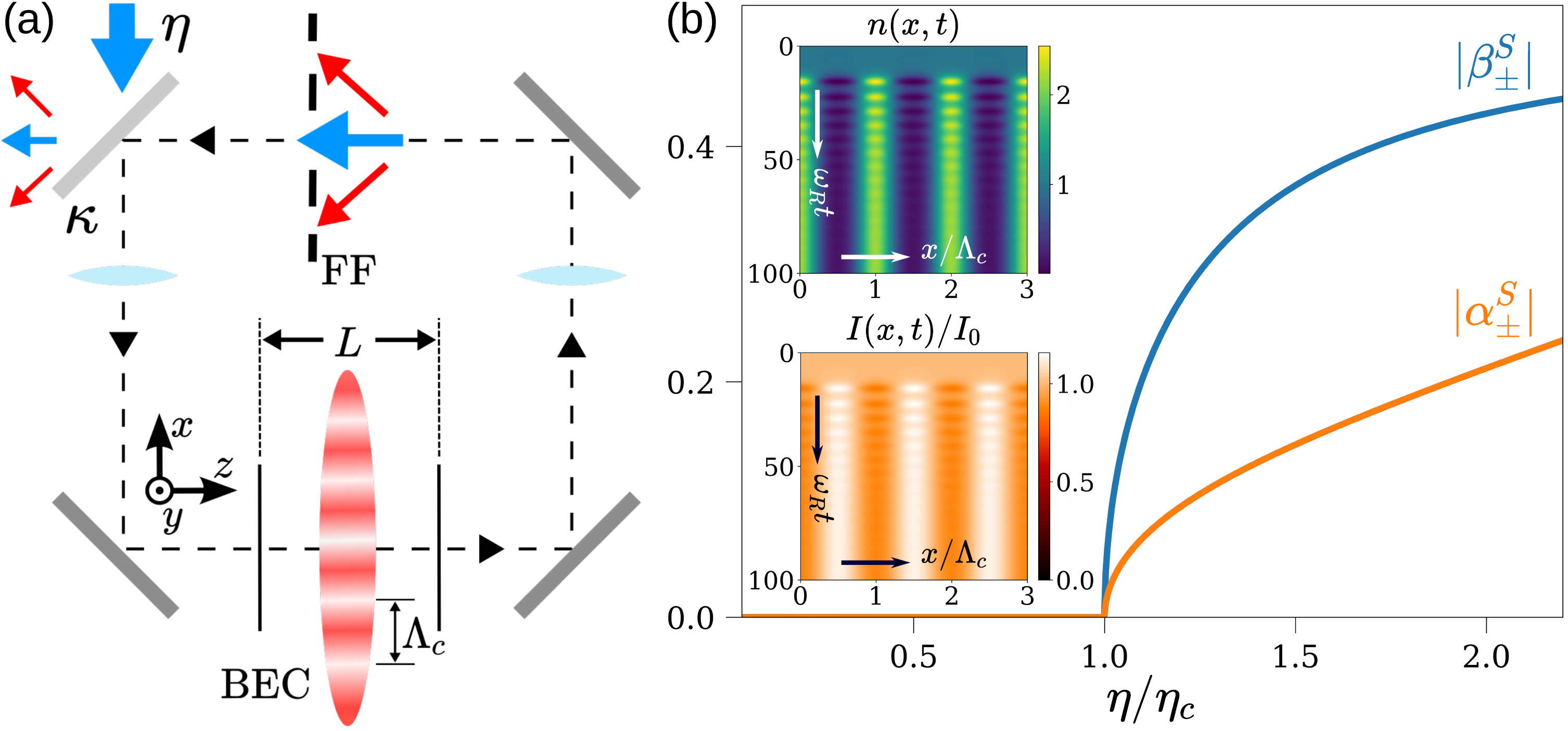}
\caption{(a) Schematics of the proposed experimental setup for observing Dicke state entanglement via self-organization of ultracold atoms in a laser pumped ring cavity. The ultracold BEC gas, strongly confined along the $y$ and $z$ axes, is placed in a ring cavity with linewidth $\kappa$, which is pumped by coherent on-axis light with drive amplitude $\eta$ (blue arrow). Effective cavity length $L$ can be controlled by adjustment of intracavity lenses (light blue) around the afocal telescopic condition, while pattern lengthscale $\Lambda_c$ is tuned via Fourier filtering (FF) of the photonic sidebands (red arrows). (b) Mean-field transverse optomechanical self-ordering in a leaky ring cavity. The square root of the steady state diffracted photon number $ |\alpha_\pm^S|$ (orange) and transverse atomic sideband population $|\beta_\pm^S|$ (blue) for varying the input beam pump rate $\eta$ (see text). Inset: temporal evolution of the semiclassical stripe patterns in the atomic density $n(x,t)$ (upper) and the normalized electric field $I(x,t)/I_0$ (lower) at $\eta = 1.2\eta_c$ (see text). Simulation parameters: $N=10^4$ and ($\bar{\Delta}_c,\;\bar{\Delta}_c',\; U_0,\;\kappa) = (8.8,\;-10,\; 1.2\times 10^{-4},\;10)\omega_R$, with $\hbar=1$.}\label{Fig:meanfield}
\end{figure}

We now numerically investigate the system dynamics in the mean-field limit of the dissipative evolution for the Hamiltonian $H_{cav}$. In Fig. \ref{Fig:meanfield}b) we plot the square root of the diffracted photon number $|\alpha_\pm^S|=(\langle n_\pm\rangle_S/N)^{1/2}$ and the transverse atomic sideband population $|\beta_\pm^S|=(\langle N_\pm\rangle_S/N)^{1/2}$ vs.\ the pump  $\eta$. For the results in Fig. \ref{Fig:meanfield}b), the equations (\ref{eq:photlang_mean})-(\ref{eq:atomheis_mean}) are solved numerically and the steady state values are plotted for different $\eta$'s. In the inset we plot the temporal evolution of the roll patterns in the atomic density, given by $n(x,t)=|\beta_0(t)+\beta_+(t)\exp(iq_cx)+\beta_-(t)\exp(-iq_cx)|^2$, and in the electric field intensity, given by $I(x,t)=|\alpha_0(t)+\alpha_+(t)\exp(iq_cx)+\alpha_-(t)\exp(-iq_cx)|^2$, normalized to the steady state $|\alpha_0^S|^2$, denoted by $I_0$.

%For $\bar{\Delta}_c< 0$, , while for $\bar{\Delta}_c> 0$ the threshold pump rate vanishes, and diverges for $\bar{\Delta}_c=0$ (see the SM for details)
The mean field patterns appear for the pump rate $\eta>\eta_c$. The initial sharp increase in the transverse excitations, seen in Fig. \ref{Fig:meanfield}b) for $\eta\gtrsim \eta_c$, gives way to saturation for larger $\eta$'s. In the inset we plot the temporal evolution of the semiclassical stripe patterns in the atomic density $n(x,t)$ and the electric field intensity $I(x,t)$, normalized to the steady state $|\alpha_0^S|^2$, denoted by $I_0$. 

The system starts with $\alpha_0(0)=\alpha_\pm(0)=0$, with $\alpha_0(t)$ rising rapidly on the scale $1/\kappa$, which is not detectable on the plot since the $t$ range is too large. This homogeneous state becomes unstable after around $10/\omega_R$, a time determined by the initial fluctuations in the atomic density, which were made artificially small in the plots shown (note that the steady state in this self-organizing mean field model is the same for all reasonable values of initial fluctuations). Before reaching the steady state, the $|\alpha_\pm(t)|$ and $|\beta_\pm(t)|$ oscillate at a frequency of a few $\omega_R$. This oscillation of the sideband populations is a signature of the sloshing dynamics, i.e.\ the continuous oscillation between the bunched and homogeneous atomic structure in the optical lattice (studied e.g. in \cite{nagy2010,tesio14}). The stripes in $n(x,t)$ and $I(x,t)$ are complementary, which is a consequence of the optical dipole potential repulsing the atoms away from the intensity peaks for $U_0>0$ \cite{labeyrie_optomechanical_2014,robb_quantum_2015}.

\section{Continuous translational symmetry of $H_B$ and $H_{cav}$}
We start by writing again the operators for the photon and atomic momentum sidebands, which have the form $\delta n=n_+-n_-$, $\delta N=N_+-N_-$, with $n_\pm=a_\pm^\dagger a_\pm$ and $N_\pm=b_\pm^\dagger b_\pm$. The Hamiltonian $H_B$ is symmetric to translations by a real-valued distance parameter $d$ along the $x$ axis, which transforms the wavefunction as $\psi(x)\to \psi(x +d)$, meaning it is symmetric under simultaneous transformations: $b_0\to b_0,\:b_\pm\to e^{\pm idk_f} b_\pm$. The generator of this symmetry is the $x$ component of the atomic momentum operator, in the many-body formalism given simply by $p_x=\hbar k_f\delta N$. The corresponding translation operation is performed by the unitary operator $T_B(d)=e^{idp_x/\hbar}$. This continuous symmetry of $H_B$ leads to:
\begin{align}\label{eq:symm_H_B}
T_B^\dagger(d) H_BT_B(d) =H_B \:\to\:[\delta N, H_B]=0.
\end{align}
Note that the same conclusion can be reached from the opposite direction, by explicitly calculating $[\delta N, H_B]=0$ and finding the corresponding unitary symmetry operator. By using the Baker-Campbell-Hausdorff formula and the commutator $[\delta N,b_\pm]=\mp b_\pm$, it can then be readily shown that $T_B^\dagger(d)b_\pm T_B(d)=e^{\pm idk_f}b_\pm$.
 
The Hamiltonian $H_{cav}$ is also symmetric to continuous translations by $d$ along the $x$ axis, which transforms the wavefunction and the electric field as $\psi(x)\to \psi(x +d)$, $E(x)\to E(x +d)$, meaning it is symmetric under simultaneous transformations: $a_0\to a_0,\:a_\pm\to e^{\pm iq_cd} a_\pm$, $b_0\to b_0,\:b_\pm\to e^{\pm iq_cd} b_\pm$. The generator of this symmetry is the $x$ component of the combined photonic and atomic momentum operator, given by $P_x=\hbar q_c(\delta n+\delta N)$. The corresponding translation operation is performed by the unitary operator $T_{cav}(d)=e^{idP_x/\hbar}$. This continuous symmetry of $H_{cav}$ leads to:
\begin{align}\label{eq:symm_H_cav}
 T_{cav}^\dagger(d)H_{cav}T_{cav}(d) =H_{cav} \:\to\:[\delta n+\delta N, H_{cav}]=0.
\end{align}
Again, the same conclusion can be reached from the opposite direction, by explicitly calculating $[\delta n+\delta N, H_{cav}]=0$ and finding the corresponding unitary symmetry operator. Following the same procedure as above, it can readily be shown that now $T_{cav}^\dagger(d)a_\pm T_{cav}(d)=e^{\pm idq_c}a_\pm$ and $T_{cav}^\dagger(d)b_\pm T_{cav}(d)=e^{\pm idq_c}b_\pm$.

To illustrate the meaning of preservation of translational symmetry during temporal evolution, we calculate the atomic spatial probability distribution $\langle \psi^\dagger (x)\psi(x)\rangle$ for $V=1$ as:
\begin{align}\label{eq:atomspatial}
\langle \psi^\dagger (x)\psi(x)\rangle=\langle N_0\rangle+\langle N_+\rangle+\langle N_-\rangle+\langle b_0^\dagger b_++b_-^\dagger b_0\rangle e^{ikx}+\langle b_0^\dagger b_-+b_+^\dagger b_0\rangle e^{-ikx}+\langle b_-^\dagger b_+\rangle e^{2ikx}+\langle b_+^\dagger b_-\rangle e^{-2ikx},
\end{align}
where $k=k_f$ for $H_B$ and $k=q_c$ for $H_{cav}$. It can readily be shown that for a translationally invariant density matrix $\rho(t)=T^\dagger(d) \rho(t)T(d)$, the inhomogeneous terms of Eq. (\ref{eq:atomspatial}) vanish (see below), and $\langle \psi^\dagger (x)\psi(x)\rangle=N$, in contrast to the results for the semiclassical (mean-field) framework shown in Fig. \ref{Fig:meanfield}. Such situation where atomic spatial probability distribution is homogeneous, even though self-organization takes place, occurs because of the Heisenberg uncertainty relation for position and momentum measurements. This means that, as the atoms are in a continuously translationally invariant state, a pattern realization with any displacement from $x=0$ (i.e. pattern phase) is equally probable.

Performing a measurement of the atomic position will collapse the system onto a state with an undetermined (i.e. one with maximum variance in) total momentum, which means the inhomogeneous terms of Eq. (\ref{eq:atomspatial}) will be nonzero. The spatial probability distribution will then be sinusoidal, with a fixed spatial phase of the patterns. As the measured states have undetermined atomic positions, the measurement of the atomic distributions should yield random values of the pattern phases. This was indeed seen in the experiments of Ref. \cite{zhang2020}, where pattern realizations with random pattern phases (displacements) and orientations, in a 2D system, were reported.

We note that the above described translationally invariant states are analogous to the maximally amplitude squeezed photonic states, for which the phase of the electric field is undetermined \cite{walls83}. Similar conclusions were previously also reached for the photonic self-organized patterns, as reported in \cite{grynberg_quantum_1993}. 

%In ultracold atom experiments, the two most common complementary measurements of density (position) and momentum distributions of atoms in a cloud are near-field microscopy and time-of-flight imaging, respectively. To efficiently generate multiparticle entangled states, the experiments should gather information by measuring the momentum distribution via quantum nondemolition measurements \cite{braginsky80}, and completely avoid measuring the density distribution, as done e.g. in the experiment of Ref. \cite{lucke14}.

\section{Vanishing of $\langle\mathbf{J}\rangle$ during dynamical evolution}
It can readily be shown that the expectation value of the total (vector) ``angular momentum" operator of the atomic momentum sidebands vanishes, i.e. $\langle\mathbf{J}\rangle=0$, for both unitary evolution under $H_B$, along with unitary and dissipative evolution under $H_{cav}$. To do this, we first write the atomic momentum ladder operators $J_\pm=b_\pm^\dagger b_\mp$, where the $J_x,\: J_y$ are now $J_x=(J_++ J_-)/2,\: J_y=(J_+-J_-)/2i$. 

For $H_B$, in our simulations the system evolves from the initial state $|\psi_0\rangle=|N\rangle_0|0\rangle_+|0\rangle_-$ to the state $|\psi(t)\rangle$, by unitary evolution $|\psi(t)\rangle = U_B(t)|\psi_0\rangle$, where $U_B(t)=e^{-iH_B t/\hbar}$. The expectation values of $J_\pm$ at time $t$ are thus given by $\langle J_\pm\rangle=\langle \psi(t)|J_\pm|\psi(t)\rangle$. Now, the temporal evolution of the state $|\psi_0\rangle$ via $H_B$ leaves the state $|\psi(t)\rangle$ on a part of the Hilbert space spanned by states with zero transverse momentum, since $H_B$ conserves the transverse momentum. In contrast, the operators $J_\pm$ act to transfer the state $|\psi(t)\rangle$ to an orthogonal part of the Hilbert space, since they increase/decrease the transverse momentum by $2\hbar k_f$. The orthogonality of $J_\pm|\psi(t)\rangle$ and $|\psi(t)\rangle$ thus leads to $\langle J_\pm\rangle=0$, and correspondingly to $\langle J_x\rangle=\langle J_y\rangle=0$. Likewise, $\langle J_z\rangle=0$ follows from the fact that the total transverse momentum of the state $|\psi(t)\rangle$ remains zero for all time.  

For unitary evolution under $H_{cav}$, the initial state is in our simulations given by $|\psi_0\rangle=|0,0,0\rangle_{ph}|N\rangle_0|0\rangle_+|0\rangle_-$. The dynamics leads now to the state $|\psi(t)\rangle = U_{cav}(t)|\psi_0\rangle$, where $U_{cav}(t)=e^{-iH_{cav} t/\hbar}$, for which the total photonic and atomic transverse momentum is equal to zero. As the operators $J_\pm$ do not conserve the transverse momentum, they take the state $|\psi(t)\rangle$ to an orthogonal part of the Hilbert space, and we again have $\langle J_x\rangle=\langle J_y\rangle=0$. 

For $H_{cav}$, $\langle J_z\rangle=0$ follows from parity ($+\leftrightarrow -$) symmetry of $H_{cav}$, i.e. $P^{-1} H_{cav} P=H_{cav}$, where the operator $P$ inverts the $x$ axis, switching between the $+$ and $-$ modes. The parity symmetry of $H_{cav}$ and $|\psi_0\rangle$ leads to a parity symmetry of the state $|\psi(t)\rangle$, which means $\langle J_z\rangle=\langle\psi(t)| J_z|\psi(t)\rangle=\langle\psi(t)|P^{-1} J_z P|\psi(t)\rangle=-\langle J_z\rangle$, leading to $\langle J_z\rangle=0$.  

To show that $\langle\mathbf{J}\rangle=0$ during dissipative evolution for $H_{cav}$, we write the Lindblad master equation for the density matrix $\rho(t)$ as \cite{breuer02}:
\begin{align}\label{eq:lindrho}
\frac{d\rho}{dt}=\mathcal{L}[\rho],\:\mbox{where}\:\: \mathcal{L}[\rho]=-\frac{i}{\hbar}[H_{cav},\rho]+\kappa\sum_{j=0,\pm}(2a_j \rho a_j^\dagger-a_j^\dagger a_j \rho-\rho a_j^\dagger a_j).
\end{align}
The above equation can be formally solved as \cite{preskill98}:
\begin{align}\label{eq:lindrhoexp}
\rho(t)=e^{\mathcal{L}t}[\rho(0)]=\rho(0)+t\mathcal{L}[\rho(0)]+\frac{t^2}{2}\mathcal{L}[\mathcal{L}[\rho(0)]]+...
\end{align}
By noting that $T_{cav}(d)\rho(0)T_{cav}^\dagger(d)=\rho(0)$, it can readily be shown that $T_{cav}(d)\rho(t)T_{cav}^\dagger(d)=\rho(t)$. To demonstrate this, we write for the first order term of (\ref{eq:lindrhoexp}):
\begin{align}\label{eq:lindrhoexp2}
T_{cav}(d)\mathcal{L}[\rho(0)]T_{cav}^\dagger(d)=T_{cav}(d)[-\frac{i}{\hbar}[H_{cav},\rho(0)]+\kappa\sum_{j=0,\pm}(2a_j \rho(0) a_j^\dagger-a_j^\dagger a_j \rho(0)-\rho(0) a_j^\dagger a_j)]T_{cav}^\dagger(d).
\end{align}
Now, the relations $T_{cav}(d)H_{cav}T_{cav}^\dagger(d)=H_{cav}$, $T_{cav}(d)a_0T_{cav}^\dagger(d)=a_0$ and $T_{cav}(d)a_\pm T_{cav}^\dagger(d)=e^{\mp i dq_c}a_\pm$ lead to:
\begin{align}\label{eq:lindrhoexp3}
T_{cav}(d)\mathcal{L}[\rho(0)]T_{cav}^\dagger(d)=\mathcal{L}[\rho(0)].
\end{align}
Similarly, we write the second order term of (\ref{eq:lindrhoexp}) as:
\begin{align}\label{eq:lindrhoexp4}
T_{cav}(d)\mathcal{L}[\mathcal{L}[\rho(0)]]T_{cav}^\dagger(d)=T_{cav}(d)[-\frac{i}{\hbar}[H_{cav},\mathcal{L}[\rho(0)]]+\kappa\sum_{j=0,\pm}(2a_j \mathcal{L}[\rho(0)] a_j^\dagger-a_j^\dagger a_j \mathcal{L}[\rho(0)]-\mathcal{L}[\rho(0)] a_j^\dagger a_j)]T_{cav}^\dagger(d).
\end{align}
Using now the same relations as above and Eq. (\ref{eq:lindrhoexp3}) leads to:
\begin{align}\label{eq:lindrhoexp5}
T_{cav}(d)\mathcal{L}[\mathcal{L}[\rho(0)]]T_{cav}^\dagger(d)=\mathcal{L}[\mathcal{L}[\rho(0)]].
\end{align}
Repeating the same procedure for all the higher order terms leads to:
\begin{align}\label{eq:lindrhoexp6}
T_{cav}(d)\rho(t)T_{cav}^\dagger(d)=\rho(t),
\end{align}
which can also be shown by using Eq. (2.4) of Ref. \cite{albert14}. 

One can then show that $\langle T_{cav}^\dagger(d)J_\pm T_{cav}(d)\rangle=e^{\mp 2idq_c}\langle J_\pm \rangle=\mbox{Tr}(T_{cav}^\dagger(d)J_\pm T_{cav}(d) \rho(t))=\mbox{Tr}(J_\pm T_{cav}(d) \rho(t)T_{cav}^\dagger(d))=\mbox{Tr}(J_\pm \rho(t))=\langle J_\pm\rangle$, where we have used the invariance of trace under cyclic permutations, and the relations above. Since $e^{\mp 2idq_c}\langle J_\pm \rangle=\langle J_\pm\rangle$ for any $d$ in the 1D space of the problem, we have $\langle J_\pm\rangle=\langle J_x\rangle=\langle J_y\rangle=0$ during dynamical evolution.

To show that $\langle J_z\rangle=0$, we first note that the same procedure as above leads from $P\rho(0)P^{-1}=\rho(0)$ to $P\rho(t)P^{-1}=\rho(t)$. We then have $\langle P^{-1}J_z P\rangle=-\langle J_z \rangle=\mbox{Tr}(P^{-1}J_z P \rho(t))=\mbox{Tr}(J_z P \rho(t)P^{-1})=\mbox{Tr}(J_z \rho(t))=\langle J_z\rangle$, which leads to $\langle J_z\rangle =0$ during dynamical evolution.

Finally, we note that for the dissipative evolution, in contrast to the unitary case, the total momentum is not conserved, even though $\langle \mathcal{L}^\dagger[\delta n+\delta N]\rangle= 0$ and $\langle\delta n\rangle=\langle\delta N\rangle=0$, as $\mathcal{L}^\dagger[\delta n+\delta N]\neq 0$ \cite{albert14}. Since $\delta n+\delta N$ does not commute with $a_\pm,\: a_\pm^\dagger$, this means that the higher moments of $\delta n+\delta N$ do not vanish in the dissipative case, see Fig. \ref{Fig:vars}a). However, the continuous translational symmetry of the state is still preserved during temporal evolution, as $\rho(t)=T_{cav}^\dagger(d) \rho(t)T_{cav}(d)$, because $T_{cav}(d) H_{cav} T_{cav}^\dagger(d)=H_{cav}$ and $T_{cav}(d) a_j T_{cav}^\dagger(d)=e^{i\phi_j}a_j$, for $j=0,\pm$. Operators of this type were discussed also in Ref. \cite{albert14}.

%More formally, it can readily be shown that $H_{cav}$ commutes with the combined difference operator for the photon intensities and the atomic momentum sideband populations, which has the form $\delta n +\delta N$, where $\delta n=n_+-n_-$, $\delta N=N_+-N_-$, with $n_\pm=a_\pm^\dagger a_\pm$ and $N_\pm=b_\pm^\dagger b_\pm$. This can be seen from $[H_0,\delta n+\delta N]=0$ and:
%\begin{align}\label{eq:commutators}
%[H_{FWM}^{(1)},n_\pm]&=\hbar U_0(-a_\pm^\dagger b_\mp^\dagger a_0b_0+a_0^\dagger b_0^\dagger a_\pm b_\mp+a_0^\dagger b_\pm^\dagger a_\pm b_0-a_\pm^\dagger b_0^\dagger a_0 b_\pm),\\
%[H_{FWM}^{(2)},n_\pm]&=\hbar U_0(\mp a_+^\dagger b_-^\dagger a_-b_+\pm a_-^\dagger b_+^\dagger a_+b_-),\\
%[H_{FWM}^{(1)},N_\pm]&=\hbar U_0(-a_\mp^\dagger b_\pm^\dagger a_0b_0+a_0^\dagger b_0^\dagger a_\mp b_\pm-a_0^\dagger b_\pm^\dagger a_\pm b_0+a_\pm^\dagger b_0^\dagger a_0 b_\pm),\\
%[H_{FWM}^{(2)},N_\pm]&=\hbar U_0(\pm a_+^\dagger b_-^\dagger a_-b_+\mp a_-^\dagger b_+^\dagger a_+b_-),
%\end{align}
%leading to $[H_{cav},\delta n+\delta N]=0$. Note that the Hamiltonian $H_{cav}$ does not commute with the individual operators $\delta n$ and $\delta N$.

%As the momentum is conserved in the photon-atoms scattering processes described by $H_{cav}$, the number of excitations is also conserved, making the Hamiltonian invariant to the transformation $(a_\pm,\:b_\pm)\leftrightarrow (b_\pm,\:a_\pm)$. Unitary evolution under $H_{cav}$

\section{Temporal evolution of the variances for $H_{cav}$}
We now discuss qualitatively the consequence of $[H_{cav},\delta n+\delta N]=0$ on the behavior of the variances of the $\delta n+\delta N$, $\delta n$ and $\delta N$ operators by looking at their temporal evolution, described by Eqs. (\ref{eq:heis1}) and (\ref{eq:lind1}). The expectation value and variance of the operator $\delta n+\delta N$ vanishes at $t=0$, as the system starts in a homogeneous state, i.e. self-organization has not yet taken place. In the unitary case, $\langle \delta n+\delta N\rangle$ remains zero for all time, as:
\begin{align}\label{eq:heis_sum}
\frac{d(\delta n+\delta N)}{dt}=\frac{i}{\hbar}[H_{cav},\delta n+\delta N]=0.
\end{align}
Using the identity $[A,BC]=[A,B]C+B[A,C]$ cyclically, the unitary temporal evolution gives also constant values for the powers of the $\delta n+\delta N$ operator at all $t$, which means that all initially vanishing moments of this operator also vanish at all $t$. If one includes the cavity photon dissipation into the picture, the Lindblad-type evolution gives:
\begin{align}\label{eq:lind_sum1}
\frac{d\langle \delta n+\delta N\rangle}{dt}=-2\kappa\langle \delta n\rangle.
\end{align}
As $\langle\delta n\rangle=0$ for all $t$ due to the parity symmetry of $\rho(t)$ (see above), the $\langle\delta n+\delta N\rangle$ will also vanish in the dissipative case. For the variance, we look at the evolution of $\langle(\delta n+\delta N)^2\rangle$, described by:
\begin{align}\label{eq:lind_sumvar}
\frac{d\langle(\delta n+\delta N)^2\rangle}{dt}=2\kappa\langle n_++n_-\rangle-4\kappa\langle(\delta n+\delta N)\delta n\rangle.
\end{align}
The right hand side no longer vanishes, which means $\langle (\delta n+\delta N)^2\rangle$ no longer vanishes for all $t$, leading to a nonvanishing variance of $\delta n+\delta N$ in the dissipative case. 

\begin{figure}[!t]
\centering
\includegraphics[clip,width=\columnwidth]{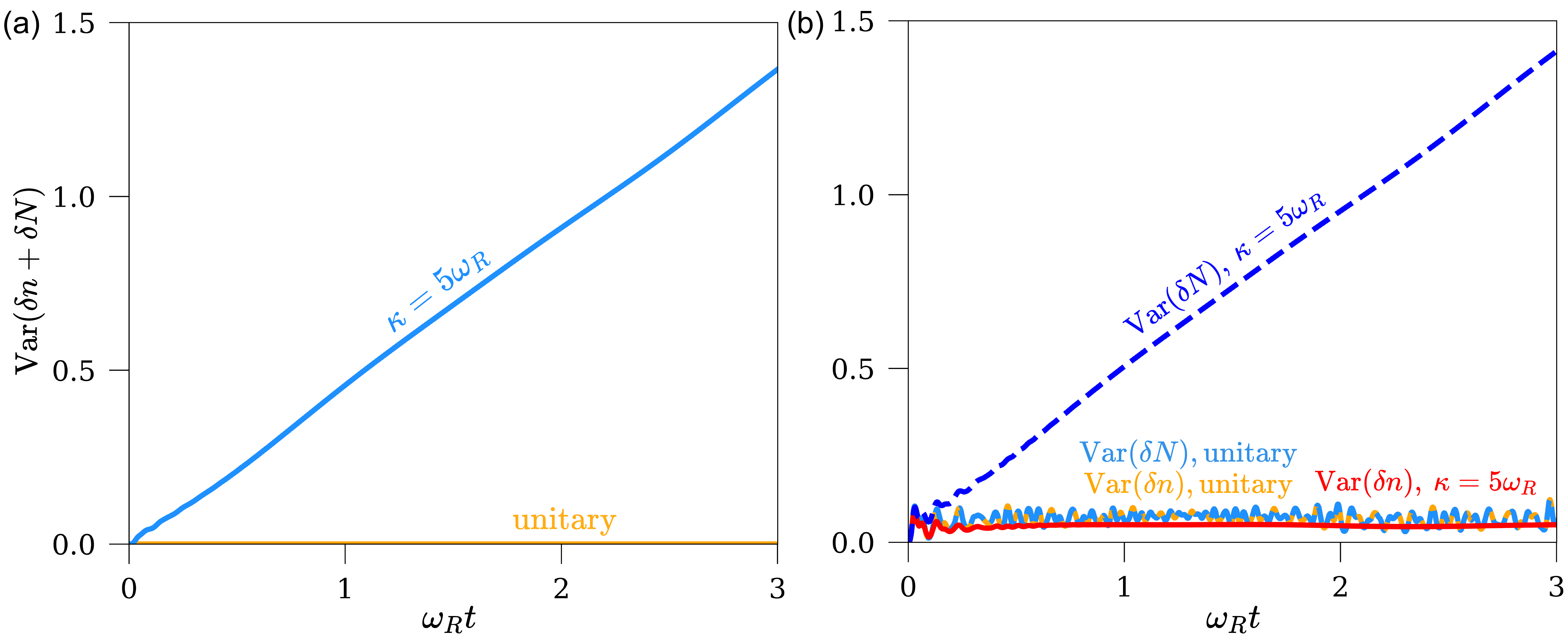}
\caption{Temporal evolution of the variances for the unitary and dissipative case with $\kappa=5\omega_R$ for the ring cavity Hamiltonian $H_{cav}$. (a) Variance of $\delta n+\delta N$ for the unitary (orange) and $\kappa=5\omega_R$ case (blue). (b) Variance of $\delta n$ for the unitary (orange) and $\kappa=5\omega_R$ (red) case, along with the variance of $\delta N$ for the unitary (light blue) and $\kappa=5\omega_R$ (dark blue) case. Simulation parameters: $N=8$, ($\eta,\;\bar{\Delta}_c,\;\bar{\Delta}_c',\; U_0) = (40,\;110,\;-45,\; 10)\omega_R$, with $\hbar=1$.}\label{Fig:vars}
\end{figure}

Fig. \ref{Fig:vars}a) shows the behavior of the $\delta n+\delta N$ variances in the case of unitary and dissipative evolution. In the unitary case, the variance vanishes (see Eq. (\ref{eq:heis_sum})), while in the dissipative case the variance increases almost linearly, indicating that the random dissipation of photons from the cavity increases the overall noise for both the photonic and atomic degrees of freedom. 

The temporal evolution of the $\delta n$ and $\delta N$ operators is correlated due to $[H_{cav},\delta n]=-[H_{cav},\delta N]$. In the unitary case the evolution of the $\delta n$ and $\delta N$ operators is related by: 
\begin{align}\label{eq:heis_deltas}
\frac{d\delta n}{dt}=\frac{i}{\hbar}[H_{cav},\delta n]=-\frac{i}{\hbar}[H_{cav},\delta N]=-\frac{d\delta N}{dt},
\end{align}
which, upon taking the expectation value and integration over $t$, leads to $\langle\delta n\rangle=-\langle \delta N\rangle$ for all $t$. Also, $\langle(\delta n)^2\rangle=\langle(\delta N)^2\rangle$ for all $t$, leading to equality of the variances of $\delta n$ and $\delta N$ for all $t$ in the unitary case (see Fig. \ref{Fig:vars}b)). Smaller $\eta$'s lead to smaller $\langle n_0\rangle$, $\langle n_\pm\rangle$ and $\delta n$ variances, meaning that the $J_z$ variance in the unitary case will be reduced for a smaller number of photons in the cavity.

For the dissipative case, the $\langle \delta n\rangle$ evolves as:
\begin{align}\label{eq:lind_deltaphotons}
\frac{d\langle\delta n\rangle}{dt}=\frac{i}{\hbar}\langle[H_{cav},\delta n]\rangle-2\kappa\langle\delta n\rangle,
\end{align}
while 
\begin{align}\label{eq:lind_deltaatoms}
\frac{d\langle\delta N\rangle}{dt}=\frac{i}{\hbar}\langle[H_{cav},\delta N]\rangle=-\frac{d\langle\delta n\rangle}{dt}-2\kappa\langle\delta n\rangle.
\end{align}
For the variances in the dissipative case we look at the evolution of $\langle(\delta n)^2\rangle$, given by:
\begin{align}\label{eq:lind_deltasqphotons}
\frac{d\langle(\delta n)^2\rangle}{dt}=\frac{i}{\hbar}\langle[H_{cav},(\delta n)^2]\rangle+2\kappa\langle n_++n_-\rangle-4\kappa\langle(\delta n)^2\rangle,
\end{align}
and the evolution of $\langle(\delta N)^2\rangle$, given by
\begin{align}\label{eq:lind_deltasqatoms}
\frac{d\langle(\delta N)^2\rangle}{dt}=\frac{i}{\hbar}\langle[H_{cav},(\delta N)^2]\rangle=-2\left\langle\left(\frac{d\delta n}{dt}+2\kappa\delta n\right)\delta N \right\rangle.
\end{align}
where we have again used $[A,BC]=[A,B]C+B[A,C]$. The variances of $\delta n$ and $\delta N$ will no longer be equal at all $t$ in the dissipative case, although the temporal evolution of these operators is still coupled.

In Fig. \ref{Fig:vars}b) we plot the evolution of $\delta n$ and $\delta N$ variances for the unitary and dissipative cases. As expected from Eq. (\ref{eq:heis_deltas}), the variances of $\delta n$ and $\delta N$ are equal for the unitary case, and not equal for the dissipative case, with the variance of $\delta N$ increasing almost linearly with time, and the $\delta n$ variance fluctuating around a small constant value. The reason for a relatively low $\delta n$ variance at the used parameters is the low intracavity photon number (see e.g. Fig. \ref{Fig:compare}), occuring due to the laser-cavity detunings $\bar{\Delta}_c,\;\bar{\Delta}_c'$ being much larger in absolute value than the cavity linewidth.

From Eq. (\ref{eq:lind_deltasqatoms}) it is also clear that when $\delta n$ is a zero matrix, which happens e.g. when there is no light in the cavity, the variance of $\delta N$ stays constant. This fact can be used to create steady state motional Dicke squeezing by turning off the pump drive at a suitable time (see Fig. \ref{Fig:entangle}).

\section{Steady state Dicke entanglement generation}
In order to generate Dicke entanglement in the steady state, we move away from continuous wave (cw) driving and instead apply a temporally tailored driving field, given by a single square pulse amplitude starting instantaneously at $t=0$ and switching off instantaneously at $t=t_{OFF}$, where $t_{OFF}$ is the time for which strongest Dicke state entanglement is observed.

\begin{figure}[!t]
\centering
\includegraphics[clip,width=1\columnwidth]{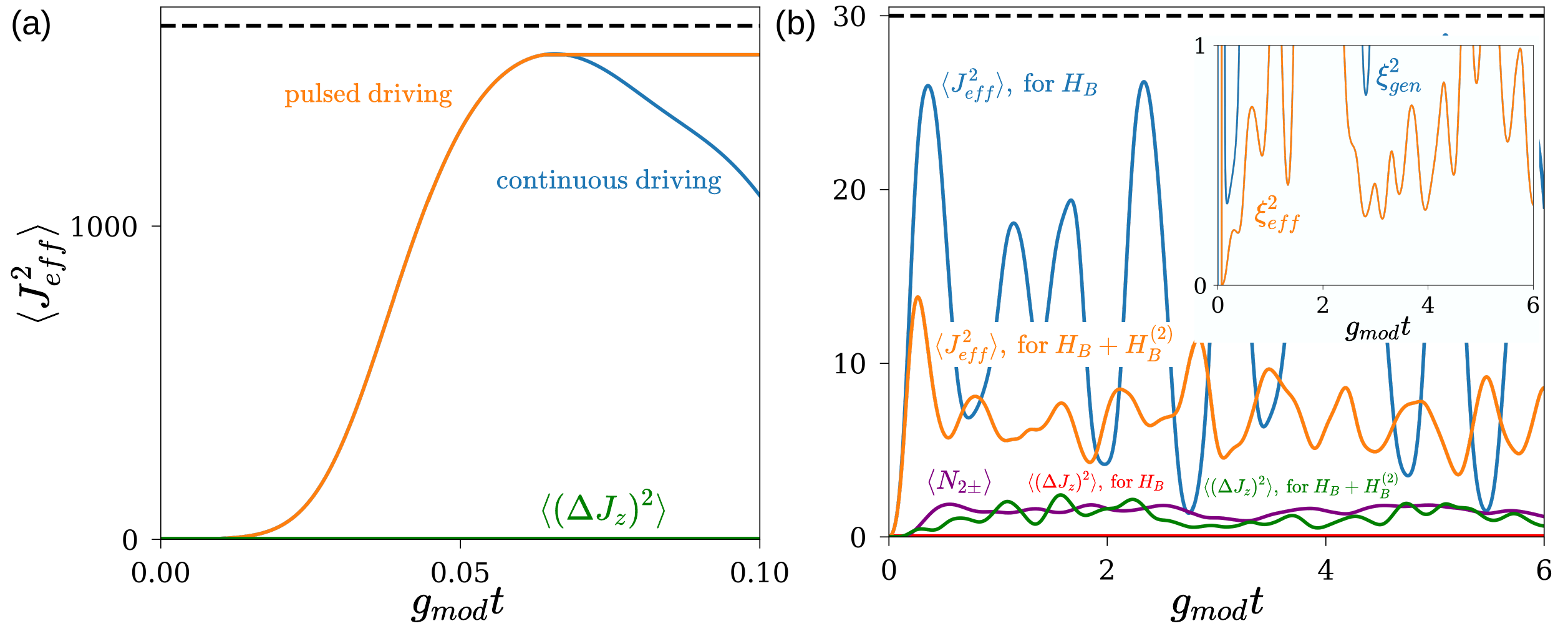}
\caption{Simulation results for unitary evolution with $H_B$ and $H_B+H_B^{(2)}$. (a) Temporal evolution of $\langle J_{eff}^2\rangle$ for pulsed driving switched off at $t_{OFF}=0.064/g_{mod}$ (orange) and continuous driving (blue) for $N=80$ atoms, see text. The green line represents $\langle (\Delta J_z)^2\rangle$ for both cases. (b) Influence of adding the second order terms $H_B^{(2)}$ on the generation of entangled states for $N=10$ atoms. Temporal evolution of $\langle J_{eff}^2\rangle$ (blue for $H_B$, orange for $H_B+H_B^{(2)}$), $\langle (\Delta J_z)^2\rangle$ (red for $H_B$, green for $H_B+H_B^{(2)}$) and $\langle N_{2\pm}\rangle$ (purple), where $N_{2\pm}=a_{2\pm}^\dagger a_{2\pm}$. Inset: Temporal evolution of $\xi_{gen}^2$ (blue) and $\xi_{eff}^2$ (orange) for the Hamiltonian $H_B+H_B^{(2)}$. Horizontal dashed black line in both plots represents the largest achievable value of $\langle J_{eff}^2\rangle$.}\label{Fig:HB}
\end{figure}

For unitary evolution with $N=80$ atoms under $H_{B}$, this switch off time is $t_{OFF}=0.064/g_{mod}$, as shown in Fig. \ref{Fig:HB}a). After switching off the driving $g_{mod}$, the $\langle J_{eff}^2\rangle$ stays near its maximum value (black dashed line) with $\langle (\Delta J_z)^2 \rangle=0$, meaning the system is very near the ideal Dicke state. Note that we have here neglected the fact that a square-shaped driving will have in the spectrum the Fourier components with frequencies not equal to $\omega_{mod}$. The undesired part of the driving spectrum can however be readily suppressed by using a suitable window function for the pulse shape. Indeed, apparently steady state sideband momentum distributions with peaks at discrete momentum states at a ring of radius $\hbar k_f$ were measured after rapidly switching off the magnetic field driving in the 2D experiment of Ref. \cite{zhang2020}.

\begin{figure}[!t]
\centering
\includegraphics[clip,width=\columnwidth]{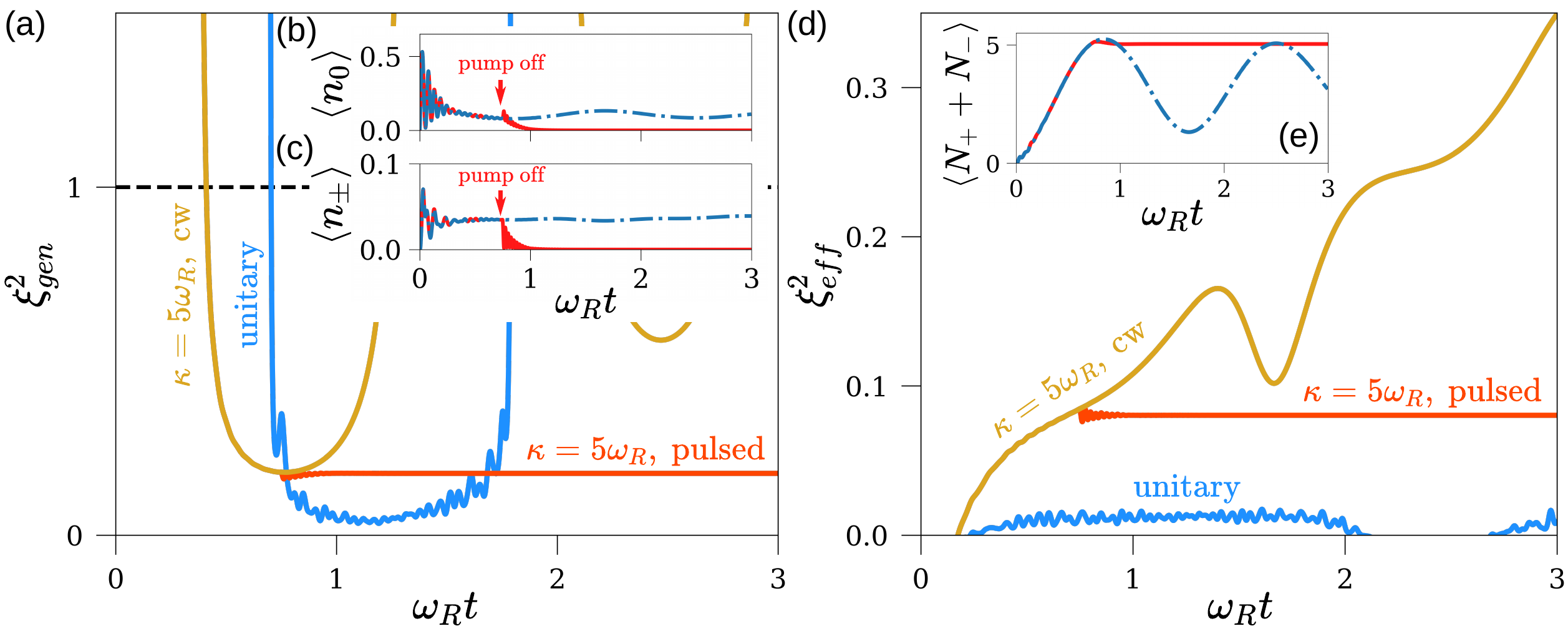}
\caption{(a) Temporal evolution of the $\xi_{gen}^2$ for the unitary (blue), dissipative cw (yellow) and pulsed (orange) cases with the largest entanglement seen in our simulations. The $\xi_{gen}^2$ reaches negative values, for which the $\xi_{gen}^2<1$ criterion is not valid, for $\langle J_{eff}^2\rangle<N/2$, so these points were excluded from the plot. The temporal evolution of (b) $\langle n_0\rangle$ and (c) $\langle n_\pm\rangle$ for the cw pump (blue, dot-dashed) and the square-pulse pump (red, solid). (d) Temporal evolution of $\xi_{eff}^2$ for the 3 cases shown in (a). (e) Temporal evolution of the number of sideband atoms for cw driving in the cw pumped case (blue, dot-dashed) and square-pulse pump (red, solid). Simulation parameters: $N=8$, ($\bar{\Delta}_c,\;\bar{\Delta}_c',\; U_0) = (110,\;-45,\; 10)\omega_R$, with $\hbar=1$. Unitary case: $\eta=40\omega_R$, dissipative case: $\eta=50\omega_R,\;\kappa=5\omega_R$. The pulse starts at $t=0$ and is turned off instantaneously at $t_{OFF}=0.75/\omega_R$.}\label{Fig:entangle}
\end{figure} 

For the dissipative evolution with $\kappa=5\omega_R$, $N=8$ atoms under $H_{cav}$, this switch off time is $t_{OFF}=0.75/\omega_R$, see Fig. \ref{Fig:entangle}a). After switching off the laser pump, the light field inside the cavity drops to zero in a time $\sim 1/\kappa$ (see Fig. \ref{Fig:entangle}b,c). In contrast, the atoms are left with a given momentum state distribution, and their kinetic energy at a constant value, as the light-matter interaction vanishes and there are no other channels for energy exchange in the system, since interatomic collisions and atom losses are neglected in our model. Indeed, $\xi_{gen}^2$ stays at a steady-state value of $0.18$, see Fig. \ref{Fig:entangle}a), which is also the lowest value attained for continuous driving. The lowest value attained in our simulations for the unitary case is $\xi_{gen}^2=0.03$.

We introduce a slightly different figure of merit $\xi_{eff}^2$, which is used in the experimental work of \cite{lucke14} for data analysis. The quantity is given by:
\begin{align}\label{eq:xi_eff}
\xi_{eff}^2=\frac{(\langle N_++N_-\rangle-1)\langle (\Delta J_z)^2\rangle}{\langle J_{eff}^2\rangle-\langle N_++N_-\rangle/2},
\end{align}
where the number of atoms in the sideband Dicke state has been postselected \cite{pezze18}, by replacing $N$ with $\langle N_++N_-\rangle$ in $\xi_{gen}^2$. As shown in Fig. Fig. \ref{Fig:entangle}e), the population of sideband atoms $\langle N_++N_-\rangle$ initially increases and then oscillates in time for cw pumping, while for the square pulse the number stays at a constant value of $\langle N_++N_-\rangle\approx 5$. For the square pulse case, the effective Dicke entanglement stays in our model at a constant value of $\xi_{eff}^2=0.08$ (-11 dB), which is comparable to -11.4 dB of Ref. \cite{lucke14}, see Fig. \ref{Fig:entangle}d).

For both models the random collisions of the atoms in a BEC were neglected. The influence of such random collisions on the lifetime of motional state multiparticle entanglement in a BEC is currently under investigation.

\section{Suppression of higher order sideband excitation}
Note that in writing the ansatz (\ref{eq:atomfield}), we have neglected the Fourier components with spatial periodicity of $\Lambda_c/2,\:\Lambda_c/3,$ ... in the atomic field, as was also done e.g. in Refs. \cite{nagy2010,baumann2010}. In a transversely pumped Fabry-Perot cavity, the occurence of these higher order components in $\psi(x)$ is suppressed by the excitation of effectively only a single longitudinal optical mode. This leads to a preference of the atoms to bunch into the optical lattice with the spatial periodicity set by the longitudinal cavity mode. Similarly, in the setup for $H_{cav}$, the suppression of higher order Fourier components of $\psi(x)$ is done by Fourier filtering of the intracavity light, which again leads to the preference for formation of an atomic field with periodicity $\Lambda_c$. Indeed, taking the self-organized potential depth in our calculations to be equal to $\hbar U_0\langle n_++n_-\rangle\approx 1\hbar\omega_R$ (see Fig. \ref{Fig:compare}i)), and the atomic kinetic energy to be $\hbar\omega_R\langle N_++N_-\rangle\approx 5\hbar\omega_R$ (see Fig. \ref{Fig:entangle}e)), the ratio of the potential depth to kinetic energy is smaller than unity, which indicates weak localization of the atoms \cite{domokos2003}. This relatively shallow potential means that the atomic field should not be deformed with respect to the shape of the self-organized optical lattice, and the Eq. (\ref{eq:atomfield}) should be a good ansatz for the atomic field operator. 

%Adding this term to the Hamiltonian
%\begin{align}\label{eq:fwmterms_3}
%H_{FWM}^{(3)}& = \hbar U_0[a_-^\dagger a_0 b_{2+}^\dagger
%b_++a_+^\dagger a_0 b_{2-}^\dagger
%b_-+a_-^\dagger a_+ b_{2+}^\dagger
%b_0+a_+^\dagger a_- b_{2-}^\dagger
%b_0+n_0(b_{2+}^\dagger b_{2+}+b_{2-}^\dagger b_{2-})+\mbox{H.c.}], 
%\end{align}
In contrast, the excitation of higher order sidebands in the experiment of Ref. \cite{zhang2020} is precluded only for $\langle N_0\rangle \gg \langle N_\pm \rangle $. To clarify this, we first write the second order atomic field operator as:
\begin{align}\label{eq:atomfield_hamb1}
\psi(\mathbf{r})=\frac{1}{\sqrt{V}}\left(b_0+b_+e^{ik_fx}+b_-e^{-ik_fx}+b_{2+}e^{2ik_fx}+b_{2-}e^{-2ik_fx}\right).
\end{align}
Integration of the corresponding interatomic scattering term of the Hamiltonian, $\int d^3r\psi(\mathbf{r})^\dagger\psi(\mathbf{r})^\dagger\psi(\mathbf{r})\psi(\mathbf{r})$, over the volume of the cloud, yields here the terms containing the second order sideband operators:
\begin{align}\label{eq:atomfield_hamb2}
& b_{2+}^\dagger b_{2+}^\dagger b_{2-} b_{2-}+b_{2-}^\dagger b_{2-}^\dagger b_{2+} b_{2+}+2b_0^\dagger b_0^\dagger b_{2+} b_{2-}+4b_{2+}^\dagger b_{2-}^\dagger b_{2+} b_{2-}
+4b_{2+}^\dagger b_{-}^\dagger b_{2+} b_{-}+4b_{0}^\dagger b_{+}^\dagger b_{2+} b_{-}\\
& +4b_{+}^\dagger b_{2+}^\dagger b_{2+} b_{+}+4b_{+}^\dagger b_{-}^\dagger b_{2+} b_{2-}
+4b_{0}^\dagger b_{2+}^\dagger b_{0} b_{2+}+2b_{+}^\dagger b_{+}^\dagger b_{0} b_{2+}
+2b_{0}^\dagger b_{2+}^\dagger b_{+} b_{+}+2b_{2+}^\dagger b_{2-}^\dagger b_{0} b_{0}\\
&+4b_{2+}^\dagger b_{2-}^\dagger b_{+} b_{-}+4b_{2+}^\dagger b_{-}^\dagger b_{+} b_{0}
+4b_{0}^\dagger b_{-}^\dagger b_+ b_{2-}+4b_{-}^\dagger b_{2-}^\dagger b_- b_{2-}
+4b_{0}^\dagger b_{2-}^\dagger b_0 b_{2-}+2b_{-}^\dagger b_{-}^\dagger b_0 b_{2-}\\
&+4b_{+}^\dagger b_{2-}^\dagger b_+ b_{2-}+2b_{0}^\dagger b_{2-}^\dagger b_- b_{-}
+4b_{+}^\dagger b_{2-}^\dagger b_0 b_{-}.
\end{align}
The dispersion for momentum states is $\varepsilon(k)=\hbar^2k^2/2m$, which gives for the single excitation states $\varepsilon_\pm=\hbar^2k_f^2/2m=\varepsilon$, while for the double excitation states one has $\varepsilon_{2\pm}=4\hbar^2k_f^2/2m=4\varepsilon$. Transforming to the rotating frame with $b_\pm\to b_\pm e^{-i\varepsilon t/\hbar}$, $b_{2\pm}\to b_{2\pm} e^{-4i\varepsilon t/\hbar}$, and keeping only the terms resonant with the driving $\pm\hbar \omega_{mod}= 2\varepsilon$, leads now to the second order Hamiltonian:
\begin{align}\label{eq:atomfield_hamb_2}
& H_B^{(2)}=i\hbar g_{mod}(b_0^\dagger b_{2+}^\dagger b_+ b_++b_0^\dagger b_{2-}^\dagger b_- b_-)+\mbox{H.c.}.
\end{align}
It is clear from the Eq. (\ref{eq:atomfield_hamb_2}) that the higher order sideband terms are small with respect to the first order terms only if the number of single sideband excitations is much smaller than the number of ground state atoms, i.e. $\langle N_0\rangle \gg \langle N_\pm \rangle $, which is satisfied for the weakly excited (collisionally thin) medium of Ref. \cite{zhang2020}.

In the results plotted in Fig. \ref{Fig:HB}b), we numerically demonstrate that the second order terms increase the $\xi_{gen}^2$ parameter for the two first order sidebands, by increasing $\langle (\Delta J_z)^2\rangle$ and reducing $\langle J_{eff}^2\rangle$. However, the $\xi_{eff}^2$ parameter still reaches quite small values (at short times after switching on the driving). We note here that it may turn out to be possible to filter out the higher order excitations in experiment, e.g. by judiciously tailoring the $a(t)$ or by removing the atoms in the second order modes from the self-organized lattice, which is a highly intriguing topic for future research, given the very large values of $\langle J_{eff}^2\rangle$ reached for the results of Fig. 2 of the main text.

\section{Experimental design of the ring cavity setup}
To estimate the attainability of pattern formation and entanglement generation for realistic parameters, we first note the experimentally available $\kappa$ values of $2\pi\times 0.13$ MHz \cite{kollar15} and $2\pi$ MHz \cite{baumann10}. Also, in state of the art experiments, cloud sizes for cigar-shaped BECs can be on the order of $\sim 250\:\mu$m \cite{nguyen2017}, which limits $\Lambda_c$ to values between approximately 1-50 $\mu$m. On the short side, $\Lambda_c$ is limited by numerical aperture of the light collection system, while on the long side it is limited by the requirement of having at least a few periods of the transverse pattern in a cloud.

\subsection{Tuning the sideband frequency $\omega_0'$ by varying effective cavity length $L$}

The cavity configuration envisaged in Fig. \ref{Fig:meanfield}a) consists of an afocal telescope in a ring cavity, such that the front focal plane of the right lens is a distance $L$ apart from the back focal plane of the left lens. This enables the tuning of the diffractive length of the cavity independent of the physical cavity length and thus enables flexibility with regard to distances of the optical elements to the BEC. In the focal plane between the two lenses, the spatial Fourier spectrum of the intra-cavity field at the position of the BEC is available for spatial filtering. The filter is transmitting the on-axis mode and the off-axis spatial sidebands at a particular transverse wavenumber $q_c$. The schematic with two intracavity lenses is conceptionally the simplest, however, to  minimize Fresnel losses the system may be implemented using curved mirrors (see below). %More details about potential experimental implementation of the schematics can be found in Refs. \cite{jensen98,ackemann2000,esteban2004}.

The sideband cavity mode wavevector $\mathbf{k}_0'$ (with length $k_0'=2\pi/\lambda_0'$) is related to the on-axis cavity mode wavevector $\mathbf{k}_0=k_0\mathbf{\hat{z}}$ (with length $k_0=2\pi/\lambda_0$) via the relation $\mathbf{k}_0'=\mathbf{k}_0+\mathbf{q}_c$, where $\mathbf{q}_c=q_c\mathbf{\hat{x}}$ (with length $q_c=2\pi/\Lambda_c$) is the transverse component of the sideband mode wavevector, which is in our case selected by Fourier filtering (see Fig. \ref{Fig:meanfield}a)). For a ring cavity without the intracavity lenses, the dispersion relation of the on-axis modes is $\omega_0=ck_0$, while for the sideband with $q_c$ it is $\omega_0'=ck_0'=c(k_0^2+q_c^2)^{1/2}=\omega_0(1+q_c^2/k_0^2)^{1/2}$.

For the cavity with two intracavity lenses of same focal length f, the diffractive length is given by $L$ (see Fig. \ref{Fig:meanfield}a), where $L$ is also the combined distance of the two lenses from the 4f configuration. In both the cavity with and without the intracavity lenses, the phase difference between the selected sideband mode and the on-axis mode after one round trip of duration $L_{cav}/c$ through the cavity is given by $\delta \phi=(\omega_0'-\omega_0)L_{cav}/c$, where $L_{cav}$ is the length of the cavity. On the other hand, the diffraction of the sideband mode through the cavity with a diffractive length $L$ leads, in the paraxial limit (valid for small $q_c/k_0'$), to a diffractive phase shift with respect to the on-axis mode of $\delta\phi\approx q_c^2L/2k_0'=q_c^2 cL/2\omega_0'$, see e.g. Eq. (2) of Ref. \cite{ackemann2021}. Equating the two expressions for the phase shift after one round trip through the cavity, the frequencies $\omega_0'$ and $\omega_0$ are in this case related via the relations (valid for small $q_c/k_0'$):
\begin{align}\label{eq:4fcavity}
\omega_0\approx\omega_0'\left(1-\frac{q_c^2c^2}{2\omega_0'^2}\frac{L}{L_{cav}} \right),\: \omega_0'\approx\frac{\omega_0}{2}\left(1+\sqrt{1+\frac{2q_c^2}{k_0^2}\frac{L}{L_{cav}}} \right).
\end{align}
The frequency $\omega_0'$ of the cavity mode with transverse wavevector $q_c$, selected by Fourier filtering, can thus be tuned relative to the on-axis mode frequency $\omega_0$, by translating the position of the intracavity lenses with respect to the 4f configuration, which changes the effective cavity length $L$. 

In the case when $L=0$ (4f condition perfectly satisfied), all sidebands have equal frequencies. This situation corresponds to the multimode degenerate cavity case, where tilted beams with $\omega=\omega_0$ are resonant to the longitudinal cavity mode, see e.g. Ref. \cite{slobodkin2022}. Near cavity degeneracy, i.e. for $L\ll L_{cav}$, the quadratic relationship $\omega_0'(q_c)-\omega_0\approx q_c^2\omega_0L/(2k_0^2L_{cav})$ \cite{ackemann2000,esteban2004} is recovered. The difference with respect to the situation studied in \cite{ackemann2000,esteban2004}, for which the patterns arise at $\Delta_{c}'=0$, is the fact that Fourier filtering in our setup allows us to select $q_c$ and therefore the sideband detuning $\bar{\Delta}_c'$ for a given pump laser frequency $\omega$. %Lastly, for $L=L_{cav}$ (no intracavity lenses), we regain the empty cavity relation by a Taylor expansion for small $q_c/k_0'$, which gives $\omega_0'\approx\omega_0(1+q_c^2/2k_0^2)$. 

We note here that the cavity mirrors might need to be slightly concave to allow intracavity propagation of sidebands with a slight tilt from the cavity axis in a system of finite extent. The stability of such cavities has been demonstrated e.g. in Refs. \cite{ackemann2000,esteban2004}, where Fabry-Perot cavities were used for experimental measurements of transverse self-organization.

\subsection{Estimating pattern formation threshold and multiparticle entanglement for realistic parameter values}
We here provide the experimental parameter values for the results of Fig. 4b) of the main text, and estimate the pattern formation threshold intensity for $N=5\times 10^5$ atoms. To start, we take the relevant on-axis mode of the ring cavity to be detuned by $\Delta_a=2\pi\times 50$ GHz from the $^{87}$Rb D2 line ($\omega_{Rb}=2\pi\times 384.23$ THz, $\lambda_{Rb}=780$ nm, transition linewidth $\Gamma=2\pi\times 6.066$ MHz, recoil frequency $\omega_r=2\pi\times 3.77$ kHz). We take the cavity length of $L_{cav}=20$ cm, the effective cavity length of $L=28.2\:\mu$m (i.e. very near the perfect 4f condition) and the free spectral range of $ c/L_{cav}= 1.5$ GHz. Note that the effective cavity length $L$ can be tuned by translating an intracavity lens, where commercially available high precision mechanical translation stages have engravings of down to 0.5 $\mu$m per division, while piezoelectric stages have even higher resolutions of down to 1 nm.

In deriving $H_{cav}$ we approximate the atom-cavity coupling $g_0$ for all three cavity modes to be equal, which is a good approximation in this case, as the mode frequencies and their volumes at the location of the atoms are approximately equal. Allowing for a reduction of the cavity finesse by the intracavity optics in the 4f configuration (``bad  cavity" regime), we take $\kappa =2\pi\times 3$ MHz (finesse of 250) and $g_0=2\pi\times 80$ kHz, and get $U_0=g_0^2/\Delta_a=2\pi\times 0.128$ Hz.

The transverse pattern recoil frequency $\omega_R$ for a sideband with $\Lambda_c=5\:\mu$m can be related to the optical transition recoil frequency $\omega_r$ via the relation $\omega_R=\omega_r\lambda_{Rb}^2/\Lambda_c^2=2\pi\times 92$ Hz, and for estimating the threshold we take the number of atoms to be $N=5\times 10^5$. Finally, we tune the laser frequency at $\bar{\Delta}_c'=-19\kappa=-2\pi\times 57$ MHz and $\bar{\Delta}_c=23\kappa=2\pi\times 69$ MHz. For this optimal squeezing case from Fig. 4b), the frequency difference of the on-axis mode and the sideband is equal to $\omega_0'-\omega_0=2\pi\times 132$ MHz. The critical pump rate is then given by $\eta_c=2\pi\times 27.6$ GHz. 

Following Ref. \cite{dombi2021}, the detuned saturation parameter of the intracavity beam is given by $s_\Delta=I/I_s/[1+(\Delta_a/\Gamma)^2]=g_0^2\langle n_0\rangle/(\Delta_a^2+\Gamma^2)$, which gives for the threshold intracavity zero-order light intensity $I_c=g_0^2N|\alpha_0^c|^2I_s/\Gamma^2=70$ mW/cm$^2$. The number of zero-order photons per atom at threshold can be estimated from Eq. (\ref{eq:alpha0}) to be equal to $|\alpha_0^c|^2=\langle n_0\rangle /N=\eta_c^2/N/(\bar{\Delta}_c^2+\kappa^2)=0.32$.

%We note at the end of this section that it may be possible to avoid the ``bad cavity" parameters in a state-of-the-art experiment, by replacing the intracavity lens-mirror pairs by suitably curved mirrors. This would reduce the scattering losses caused by finite reflectivity of the lens elements, and thus lead to an increase of the cavity finesse as compared to the system with intracavity lenses, which should benefit both the reduction of $\kappa$ and the increase of $g_0$. There will be other experimental challenges which will need consideration and engineering efforts, such as the losses at the Fourier filtering stage and the finite extension of the medium and beam\textbf{, however experimental realization of significant momentum entanglement should be within reach by using existing optomechanical components in a cavity built externally to the vacuum chamber of the BEC.}

\section{Adiabatic elimination of photonic modes for $H_{cav}$}
To elucidate the relationship between $H_B$ and $H_{cav}$, we adiabatically eliminate the photonic degrees of freedom, following the approach of \cite{finger2023}. We start with the Hamiltonian:  
\begin{align}\label{eq:fwmagain}
 \frac{H_{cav}'}{\hbar} =-\bar{\Delta}_c n_0-\bar{\Delta}_c'(n_++n_-)+ \omega_R(N_++N_-)  + U_0 [( b_-^\dagger a_+^\dagger + b_+^\dagger a_-^\dagger )a_0 b_0 +a_0^\dagger(b_+^\dagger a_+ +b_-^\dagger a_-)b_0 +\mbox{H.c.}], 
\end{align}
where the saturation four-wave mixing terms were neglected. The $a_0$ evolution is now given by the input-output equation \cite{gardiner1985,schleiersmith2010}:
\begin{align}\label{eq:phot0_2}
 \dot{a}_0 &=(i\bar{\Delta}_c-\kappa) a_0 -iU_0[(b_+^\dagger a_++b_-^\dagger a_-)b_0+b_0^\dagger(a_+b_-+a_-b_+)] +\sqrt{\kappa}(c_1+c_2),
\end{align}
where $c_{1,2}$ are the cavity input field operators from left (clockwise) and right (counterclokwise) directions. For a coherent input state from the counterclockwise direction, given by $|\eta '\rangle$, we have $c_1|\eta '\rangle=0$ and $c_2|\eta '\rangle=\sqrt{\kappa}\eta '|\eta '\rangle$. The term proportional to $U_0$ in Eq. (\ref{eq:phot0_2}) describes the influence of scattering into/out of the sideband photonic modes on the phase and amplitude of the on-axis field. This term can be neglected near threshold, where the on-axis mode is strongly populated compared to the sidebands. This results in the steady state solution:
\begin{align}\label{eq:phot0adiab}
a_0|\eta'\rangle=\frac{\kappa\eta'}{-i\bar{\Delta}_c+\kappa}|\eta'\rangle,
\end{align}
Tracing over the zero-order photonic subspace for the input state $|\eta '\rangle$, one gets $a_0\to \langle \eta'| a_0|\eta' \rangle=\sqrt{N}\alpha_0^{eff}$, where 
\begin{align}\label{eq:alpha0eff}
\alpha_0^{eff}=\frac{1}{\sqrt{N}}\frac{\eta_{eff}}{-i\bar{\Delta}_c+\kappa}.
\end{align}
For simplicity the trace notation is omitted in the calculations below, but it is always implied. Neglecting of the saturation four-wave mixing terms, and on-axis photonic mode depletion by scattering into sidebands, will lead to the rescaling of the pump rate $\eta_{eff}$ with respect to the pump rate $\eta$ of the full model. For the range of parameters discussed in the main text, it turns out that this can be taken into account by a single scaling factor $\eta_{eff}= 0.56\eta$, as displayed in Fig. 3c).

The operator evolution equations for $a_\pm$ are now:
\begin{align}\label{eq:photpm_2}
 \dot{a}_\pm &=(i\bar{\Delta}_c'-\kappa) a_\pm -iU_0\sqrt{N}\alpha_0^{eff} (b_\mp^\dagger b_0+b_0^\dagger b_\pm).
\end{align}
Note that, as in \cite{finger2023}, the photonic sideband Langevin noise terms are neglected in Eqs. (\ref{eq:phot0_2}, \ref{eq:photpm_2}). This assumes that the photon quantum noise has a negligible influence on the atomic motion. The atomic momentum fluctuations, induced by photons leaking out of the cavity without producing atom momentum pairs, will be taken into account via the Lindblad master equation (see below).

The coupling to the environment is modeled by the Lindblad Eq. (\ref{eq:lindrho}), which preserves translational symmetry [see Eq. (\ref{eq:lindrhoexp6})], meaning that the atomic density distribution will be homogeneous during irreversible evolution. Measurement-induced effects, which in single mode cavities can lead to atomic momentum diffusion near cavity resonance \cite{murch08,nagy09}, will be studied in future work. 

%Moreover, since the Lindblad Eq. (\ref{eq:lindrho}) describes a situation where only the photon number (and not phase) measurements are effectively performed, the evolution will preserve translational symmetry [see Eq. (\ref{eq:lindrhoexp6})], meaning that the atomic density distribution will be homogeneous. The corresponding measurement-induced collapse of the atomic wavefunction, which in single mode cavities would lead to atomic momentum diffusion \cite{murch08,nagy09}, does not occur for photon number measurements in our continuously translationally symmetric case, and neglecting the backaction induced by the Langevin noise is justified. 

Assuming that the fast photonic degrees of freedom adiabatically follow the slow atomic motional degrees of freedom (valid for $|\bar{\Delta}_c|,\:|\bar{\Delta}_c'|\gg \omega_R$), now gives [putting $\dot{a}_\pm=0$ in Eq. (\ref{eq:photpm_2})]:
\begin{align}\label{eq:photpm_adijab}
a_\pm &= \frac{iU_0\sqrt{N}\alpha_0^{eff}}{i\bar{\Delta}_c'-\kappa}(b_\mp^\dagger b_0+b_0^\dagger b_\pm).
\end{align}
Putting these terms into the Eq. (\ref{eq:fwmagain}), one gets the Hamiltonian $H_{ad}=H_{pair}+H_{ad}'$, where:
\begin{align}\label{eq:fwmadiabat}
H_{pair}= -\hbar g_{cav} (b_+^\dagger b_-^\dagger b_0b_0+b_0^\dagger b_0^\dagger b_+b_-),\: H_{ad}' =  \hbar\omega_R\left\{(N_++N_-)- \left[(2N_0-1)(N_++N_-)-2N_0\right]\frac{1}{4N}\frac{\eta_{eff}^2}{\eta_c^2} \right\} , 
\end{align}
with:
\begin{align}\label{eq:coupcav}
 g_{cav} =  \frac{2U_0^2\eta_{eff}^2|\bar{\Delta}_c'|}{(\bar{\Delta}_c'^2+\kappa^2)(\bar{\Delta}_c^2+\kappa^2)}=\frac{\omega_R }{2N}\frac{\eta_{eff}^2}{\eta_c^2} , 
\end{align}
where $\bar{\Delta}_c'<0$, as in the remainder of the manuscript. The $H_{ad}'$ term describes the $\eta_{eff}$ dependent relative energy shifts of the three momentum states, which give the energy cost or gain of pattern formation. The momentum mixing term $H_{pair}$ is proportional to $H_B$ for driving with a cosine function [see Eq. (\ref{eq:eqhambfinalprime})]. 

To include fluctuations due to photons decaying out of the cavity, we use the Linbdblad master equation:
\begin{align}\label{eq:lindrhoadiab}
\frac{d\rho}{dt}=-\frac{i}{\hbar}[H_{cav},\rho]+\kappa\sum_{j=0,\pm}(2a_j \rho a_j^\dagger-a_j^\dagger a_j \rho-\rho a_j^\dagger a_j). 
\end{align}
Taking the above mentioned tracing over the coherent input photonic state, and adiabatically eliminating the sideband photonic operators, one gets the master equation describing the evolution of the atomic momentum density matrix $\rho_{at}$:
\begin{align}\label{eq:lindrhoadiab2}
\frac{d\rho_{at}}{dt}=-\frac{i}{\hbar}[H_{ad},\rho_{at}]+\gamma\sum_{j=\pm}(2K_j \rho_{at} K_j^\dagger-K_j^\dagger K_j \rho_{at}-\rho_{at} K_j^\dagger K_j), 
\end{align}
with:
\begin{align}\label{eq:newjump}
\gamma = \frac{ U_0^2 \eta_{eff}^2\kappa}{(\bar{\Delta}_c'^2+\kappa^2)(\bar{\Delta}_c^2+\kappa^2)},\;K_\pm=(b_\mp^\dagger b_0+b_0^\dagger b_\pm). 
\end{align}

The derivation of $H_{ad}$ can shed some light on the physical mechanism leading to the production of correlated pairs via $H_{pair}$. In $H_{cav}'$, the term $h_1= \hbar U_0\sqrt{N}\alpha_0^{eff} ( b_-^\dagger a_+^\dagger + b_+^\dagger a_-^\dagger ) b_0$, describes the excitation of $\pm$ atomic sidebands by scattering from the on-axis mode into the $\mp$ photonic mode, while the term $h_2= \hbar U_0 \sqrt{N}\alpha_0^{eff *}(b_+^\dagger a_+ +b_-^\dagger a_-)b_0$ describes excitation of $\pm$ atomic sidebands by scattering from the $\pm$ photonic mode into the on-axis mode. Using $h_1$ to derive the dynamical equations for $a_\pm$, one gets: $a_\pm^{(1)}\propto b_\mp^\dag b_0$, while using $h_2$ to derive the dynamical equations for $a_\pm^\dagger$, one gets: $a_\pm^{(2)\dag}\propto b_\pm^\dag b_0$, after adiabatic elimination.

Inserting now $a_\pm^{(1)}$ into $h_2$, and $a_\pm^{(2)\dag}$ into $h_1$, one gets the terms proportional to $b_+^\dagger b_-^\dagger b_0 b_0$ from the interaction part of $H_{cav}'$. The photon associated with the process $h_1$ ($h_2$) creates correlated pairs of atomic momentum sidebands when scattering from the atoms via process $h_2$ ($h_1$). The creation of the atomic momentum pairs thus relies on photons associated with both $h_1$ and $h_2$, and not the photons associated with only $h_1$ or only $h_2$. The term $b_0^\dagger b_0^\dagger b_+ b_-$ in the interaction part of $H_{cav}'$, is derived by considering the conjugate process, governed by $h_1^\dagger$ and $h_2^\dagger$. The terms in $H_{ad}'$ are derived by inserting $a_\pm^{(1)}$ ($a_\pm^{(1)\dagger}$) into $h_1^\dagger$ ($h_1$), and $a_\pm^{(2)}$ ($a_\pm^{(2)\dagger}$) into $h_2$ ($h_2^\dagger$).

\section{Limitations for the numerical simulations with $H_{cav}$}
When solving the Schrödinger and master equations for $H_{cav}$ numerically, we truncate the infinite-dimensional Fock space of the photonic degrees of freedom into a finite-sized Fock space. This limits the maximal pump strength $\eta$ that can be used in our simulations, as higher pump rates will naturally lead to larger $\langle n_0\rangle$ and $\langle n_\pm \rangle$ values, such that higher dimensional photonic Fock spaces are needed to correctly capture the system dynamics. The limits of the calculations are in our case set by the size of the working memory of the computational nodes.

\begin{figure}[!t]
\centering
\includegraphics[clip,width=\columnwidth]{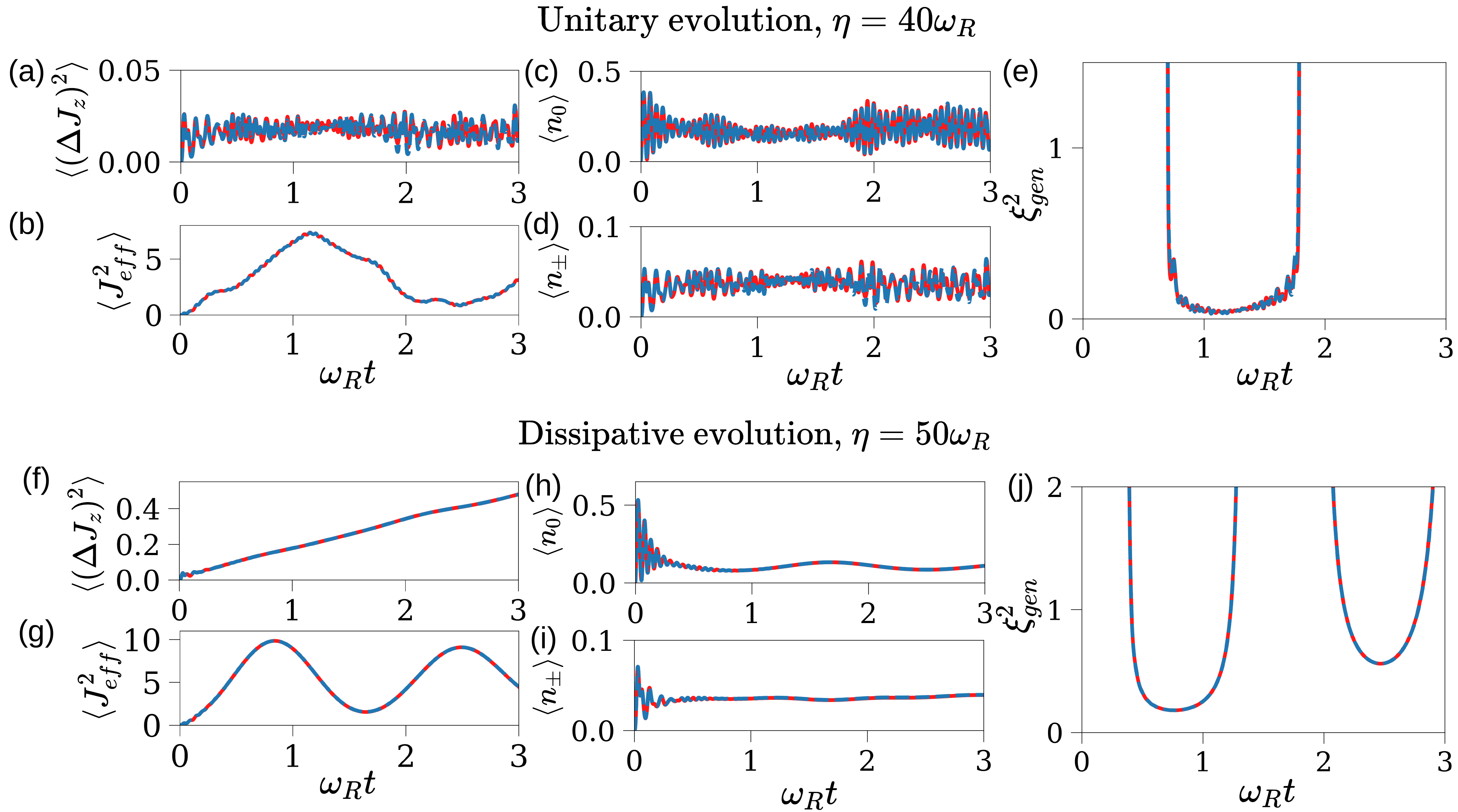}
\caption{Comparison of the temporal evolution under $H_{cav}$ of relevant system observables for the case when the state with maximal photon number in all modes is $|4\rangle_0|3\rangle_+|3\rangle_-$ (red, solid) and the case when the state with maximal photon number in all modes is $|5\rangle_0|3\rangle_+|2\rangle_-$ (blue, dashed), see text for details. The evolution of observables for the unitary case with $\eta=40\omega_R$: (a) $\langle (\Delta J_z)^2\rangle$, (b) $\langle J_{eff}^2\rangle$, (c) $\langle n_0\rangle$, (d) $\langle n_\pm\rangle$ and (e) $\xi_{gen}^2$. The evolution of observables for the dissipative case with $\eta=50\omega_R$: (f) $\langle (\Delta J_z)^2\rangle$, (g) $\langle J_{eff}^2\rangle$, (h) $\langle n_0\rangle$, (i) $\langle n_\pm\rangle$ and (j) $\xi_{gen}^2$. Simulation parameters: $N=8$, ($\bar{\Delta}_c,\;\bar{\Delta}_c',\; U_0,\;\kappa) = (110,\;-45,\; 10,\;5)\omega_R$, with $\hbar=1$.}\label{Fig:compare}
\end{figure} 

To demonstrate the numerical accuracy of the plots in the main text, we have compared the results for the dynamical evolution of the relevant observables for the case where the maximal photon number state in all modes is $|4\rangle_0|3\rangle_+|3\rangle_-$, to the case with the same total maximal photon number but where the state with the maximal photon number in all modes is $|5\rangle_0|3\rangle_+|2\rangle_-$, both at $N=8$ atoms, and at largest $\eta$ used in the unitary and dissipative cases.

In Fig. \ref{Fig:compare} we compare the two Fock space truncations by plotting the temporal evolution of relevant variables for the maximum $\eta$ values used in the unitary ($\eta=40\omega_R$) and dissipative ($\eta=50\omega_R$) cases of the main text. The plots confirm that the truncation of Fock space where the state with the maximal number of photons in each mode is $|4\rangle_0|3\rangle_+|3\rangle_-$, accurately describes the dynamics in the simulated time interval.

In the unitary case the evolution of both the atomic and photonic variables exhibit fast oscillatory motion, a signature of Vacuum Rabi oscillations occuring in the quantum electrodynamic treatment of light-matter interaction of atoms in a cavity \cite{shore93}. In the dissipative case with $\kappa= 5$, these fast oscillations are averaged out. However, the slow oscillations, a signature of the sloshing dynamics where atoms slosh around the minima of the dynamical potential and periodically amplify the transverse patterns \cite{tesio14}, still persist.
%\section{Estimation of Fock space truncation error}